\documentstyle[12pt]{article}

\begin{document}
\hfill CPTH-S590.1297\\
\hfill hep-th/9712153\\
\begin{center}

{\Large {\bf {\sf Implementation of an iterative map in the construction
of (quasi)periodic instantons: chaotic aspects and discontinuous
rotation numbers.}}}
\\[2cm]

\smallskip

{\large A. Chakrabarti\footnote{chakra@pth.polytechnique.fr}}

\smallskip

\smallskip

{\em \footnote{Laboratoire Propre du CNRS UPR A.0014}Centre de Physique
Th\' eorique, Ecole Polytechnique
 
91128 Palaiseau, France.}

\end{center}

\vfill

\begin{abstract}

An iterative map of the unit disc in the complex plane (Appendix) is used
to explore certain aspects of selfdual, four dimensional gauge fields
(quasi)periodic in the Euclidean time. These fields are characterized by
two topological numbers and contain standard instantons and monopoles as
different limits. The iterations do not correspond directly to a
discretized time evolution of the gauge fields. They are implemented in 
an indirect fashion. First, $(t,r,\theta,\phi)$ being the standard
coordinates, the $(r,t)$ half plane is mapped on the unit disc in an
appropriate way. This provides an $(r,t)$ parametrization (Sec.1) of
$Z_0$, the starting point of the iterations and makes the iterates
increasingly complex functions of $r$ and $t$. These are then
incorporated as building blocks in the generating function of the fields
(Sec.2). We explain (starting in Sec.1 and at  different stages) in what
sense and to what extent some remarkable features of our map (indicated
in the title) are thus carried over into the {\it continuous} time
development of the fields. Special features for quasiperiodicity are
studied (Sec.3). Spinor solutions (Sec.4) and propagators (Sec.5) are
discussed from the point of view of the mapping. Several possible
generalizations are indicated (Sec.6). Some broader topics are discussd
in conclusion (Sec.7).

\end{abstract}
\vfill
\vfill

\newpage
\pagestyle{plain}

\section{ Introduction}

   An iterative map of the unit disc centered at the origin of the
complex plane is studied in the Appendix(App). The motivation is that
it can be implemented  fruitfully in the study of a hierarchy of
fourdimensional, selfdual, (quasi) periodic gauge fields. Fields with
spherical symmetry in $R^{3}$ are mostly used to illustrate our approach.
More general possibilities are indicated at the end(Sec.6).Such
fields have been studied previously in a series of papers$\lbrack
1,2,3,4,5,6,7,8
\rbrack$ which contains references to other sources. Here we reformulate
them from the point of view of the iterative map.This brings remarkable
new aspects to light.
  
   Our gauge fields are are (quasi) periodic in Euclidean time. The basic
ingredient ( one may say the generating function ) for the spherically
symmetric fields is a holomorphic function $g(r+it)$ satisfying certain
constraints (Sec.2) in the $(r,t)$ half plane $(r\geq 0)$. Here      
$(t,r,\theta,\phi)$ are the standard coordinates.The function $g$ can
have several factors ( or nonfactorized terms) each with its own period
in $t$. When all the periods are mutually commensurable the lowest common
multiple of the component periods is the overall one. This is the
periodic case. When at least one of the component periods is
incommensurable with some others, one has , by definition
quasiperiodicity.

   As is well-known, fields periodic in theEuclidean time provide one
possible approach to field theory at finite temperature. Our explicit
constructions are such that even an infinitesimal change in one single
parameter (determining one of the component periods) can make a periodic
solution quasiperiodic and vice versa. So they are considered in a
parallel fashion.

     The map studied in App is
\begin{eqnarray}
     Z_{p+1} &
= &{{{a_{p}+Z_{p}}}\over{\overline{a}_{p}+{Z_{p}}^{-1}}},\qquad
\mid Z_{0}\mid\leq1;0<\mid a_{p}\mid <1 
\end{eqnarray} 
Note the inverse $({Z_{p}}^{-1})$ in the denominator. Possible choices
of $a_{p}$ are discussed in App.
{\it   For implementing the map in the construction of the gauge fields
the crucial step is a suitable $(r,t)$ parametrization of $Z_{0}$.} The
two choices considered are
\begin{eqnarray}
   Z_{0} = e^{-k(r+it)}\quad\left(k>0)\right.
\end{eqnarray}
and
\begin{eqnarray} 
   Z_{0} =  {{{\sum_{l=1}^{n}\lambda_{l}^{2}}}(
e^{k_{l}(r+it-ic_{l})}-1)^{-1}\over
   {\sum_{l=1}^{n}\lambda_{l}^{2}({1-
e^{-k_{l}(r+it-ic_{l})})^{-1}}}} 
 \end{eqnarray}              
with real parameters $(\lambda,k,c)$ and 
        $$ k_n > k_{(n-1)} > ......> k_2 > k_1> 0.$$
The uses of (2) will be amply illustrated (Secs.2,3 and App.). The
choice (3) is particularly suited to the construction of spinor solutions
(Sec.4). It can be shown to satisfy all the necessary constraints.
  
        With $Z_{0}$ thus chosen one can set (Sec.2) the generating
function of the gauge fields to be
 \begin{eqnarray}  
  g =  \Biggl(\displaystyle \prod_{j=1}^{n}Z_{p_j}^{(j)}\Biggr) 
\end{eqnarray}
where different sets of parameters are associated to each factor.

   {\it The iterations do not correspond to a discretized time
evolution of the fields}. At each step one has a different
action ( Sec.2), a different system. But the $(r,t)$ parametrization of 
$Z_{0}^{(j)}$ will provide the key to the usefulness of the iterations.
 As $p_{j}$ increases $g$ becomes a more and more complicated function
of $(r,t)$. But the fact that such complications are introduced in a
very specific fashion stepwise makes some remarkable properties
readily accessible. Some crucial properties of each block
$Z_{p_j}^{(j)}$ as a whole are delivered directly. To give this
statement explicit content let us look at the two most striking
features of our map.( Upto Sec.4, (2) will usually be referred to
directly for simplicity. This is not an essential restriction.)

  \underline{(1)}. For $r = 0$ (the circumference of the unit disc) $Z_p$
for any p, is a phase, the angle being denoted by $\psi_p$, and (1)
becomes a circle map. This map (App) satisfies all the criteria for
being chaotic ( [9], Def.8.5, p.50). These are the following ones.

  (a) A sensitive dependence on initial data, encoded by a positive
characteristic index.

  (b) A dense set of periodic points.
  
  (c) Transitivity.

  Moreover this phase and its derivatives provide the coefficients of a
series expansion in $r$ of $Z$ or $g$ (App).

       Suppose now, that as a consequence of (a), namely the positive
index, at a certain level $l$ of iterations the difference              
            $$\psi_l(\psi_0 + \delta\psi_0) - \psi_l(\psi_0)$$
is appreciably large even for a small $\delta\psi_0$ (considering
$\psi_l$ as a function of its initial datum $\psi_0$). Due to our
parametrization this means
            $$\psi_l(t + \delta t) - \psi_l(t)$$ 
is large even for small $\delta t$.

  Through (4) the result above, for each factor $j$, is injected into the
time evolution of the fields. {\it This leads to sensitive time
dependence.}( See however the remarks in Sec.7.)

         The overall phase of $g$ (for any $r$) does not contribute
directly to the action density, expressible in terms of $g\overline g$
and its derivatives (Sec.2). But the amplitude at each order of
iteration involves the phases of the lower ones (App). Moreover the
phase{ \it is }directly involved in the power series expansion in $r$.

  The phase can be studied directly and analogously for any value of $r$ 
(inside the disc). The emphasis on the phase at $r = 0$ is due to two
reasons.

(a) The relative simplicity for $r = 0$ permits a transparent derivation
of the crucial properties.

(b) For spherical symmetry, in our ansatz, the time dependence is
damped exponentially with increasing $r$ ( going toward the centre of
the disc). Hence a small sphere around the origin in $R^3$ is the most
suitable domain for studying the time evolution. The leading term of
the action density for small $r$ is given explicitly (Sec.2) to display
the role of the iterated phases.

 \underline{(2)}. We do not ignore however other remarkable features
associated with certain values of $r$ away from the spatial origine. (
The choice of the parameters $a$ and $k$ determines how far away or how
close.) These are the domains on which the $Z^{(j)}$ 's can vanish and
hence also $g$. They lead to discontinuous rotation numbers associated
to the phases (App). So far as the zeros of $g$ can be located their
cumulative effects lead to  staircase like patterns. An example is given
in Sec.2.
 
    The rotation numbers are {\it asymptotic} quantities ($n
\rightarrow \infty $ in (A58)). Hence discontinuities can arise even if
we are dealing with well-behaved, integrable action densities.

    In [10] it is emphasized ( p.20), in the context of standard circle
maps, that {\it discontinuous} rotation numbers can arise for {\it
smooth} maps. Here they arise in the context of "annular" maps (App)
when the annulus can become a disc. It is also closely related to the
central property of our map that after each iteration  "on the average"
the phase turns twice as fast (App).

     In Sec.2 the construction of the periodic fields is reformulated,
with respect to our previous papers, in order to display prominently
the role of the iterative map. The selfdual solutions are characterized
by {\it two} topological integers. One is "instanton-like" $(P_{T})$
and the other is "monopole-like" $(q)$. Their remarkable combined role 
in index theorems with $R^3 \times S^1$ as the base-manifold for
periodic fields [2,4,6,7] are recapitulated in the context of explicit
construction of spinor solutions (Sec.4).

      Some special features of quasiperiodic fields are studied in
Sec.3. For periodic backgrounds a finite number of normalizable,
zero-mass spinor solutions are obtained by imposing on them
(anti)periodic boundary conditions in $t$ (Sec.4). For quasiperiodic
backgrounds the number cannot thus be limited, unless rational
approximations of the component periods are considered. Nevertheless
the spinor solutions are constructed in a way that works for both
classes. The aim ( not realized here ) is to study the effect on the
spinors of {\it several} mutually incommensurable periods in the
background field.

      For some particularly simple systems the possible effects of
quasiperiodic kicks have been studied by several authors with
different conclusions. ( See [11] and sources cited there.)

     In a more general context ( quasiperiodic state with $k$
frequencies and a quasiperiodic attractor for the value $\mu = \mu_0$ 
of a continuous bifurcation parameter ) the situation has been summed
up as follows ([12], p.631) 

     "For $k\geq3$, strange attractors and positive characteristic
exponent may be present for $\mu$ arbitrarily close to $\mu_0$."

     We provide ( though only for zero mass and Euclidean signature )
exact, explicit solutions for spinors in four dimensions in a gauge
field background that can bring into play an arbitrary number of
mutually incommensurable periods. This can provide an interesting
starting point for further investigations.

     Our iterations can also be implemented in propagators (Sec.5).
This can lead to a systematic semiclassical development for our
classes of background.

      Several directions are indicated (Sec.6) for possible
generalizations of our study. The possibilities mentioned are --
breaking spherical symmetry, magnetic charge $q\geq1$, gauge group
$SU(N)$ with $N > 2$ and the use of hyperbolic coordinates.

      After presenting our formalism in full, certain general questions
are taken up in the concluding remarks (Sec.7).

 \section {A class of periodic selfdual gauge fields:}
  
 The class we consider, to start with, has spherical symmetry in $R^3$
and periodicity in Euclidean time. The gauge group is $SU(2)$. Let $(t,r,
\theta, \phi)$ be the standard time and radial coordinates with
 $$ds^2 = dt^2 + dr^2 + r^{2}(d\theta^2 + (sin\theta)^2  d\phi^2)$$
Let $(\sigma_r, \sigma_\theta, \sigma_\phi)$ denote the projections of
the Pauli matrices respectively on the directions indicated. The gauge
potentials are given by
\begin{eqnarray}
A_{r} \pm iA_{t} = \pm  i(\partial_r \pm i\partial_t)\zeta
\frac{\sigma_r}{2}
\end{eqnarray}
\begin{eqnarray}
A_\theta \pm i(sin \theta)^{-1}A_\phi = \pm  i(e^\zeta - 1)(
\frac{\sigma_\theta \pm i\sigma_\phi}{2}) 
 \end{eqnarray}
where
\begin{eqnarray}
e^\zeta = \frac{r}{(1 - g\bar{g})} \Bigl( ({\partial_r}^2 + 
{\partial_t}^2)(g\bar{g})\Bigr)^{ \frac{1}{2}}
\end{eqnarray}
Here $g$ is a holomorphic function $g(r + it)$ in the $(r,t)$
half-plane $(r \geq0)$, postulated to satify the following properties:

(1)  $g\bar{g} = 1 + O(r)$ for $r \rightarrow0$

(2)  $g$ has no poles for $r \geq0$

(3) $g$ falls exponentially as $r \rightarrow\infty$  (giving a   
constant logarithmic derivative)

(4) $g$ is periodic in $t$.

We start with strict periodicity. Quasiperiodicity will be defined and
studied in Sec.3. Our previous studies of periodic instantons ([1]Êto
[8]) will be reformulated below to display at each stage the role of the
iterative map (App).

\subsection{ First iteration: (from monopoles to periodic instantons)}

The simplest choice satisfying all the constraints is evidently
\begin{eqnarray}
   g = e^{-k(r+it)}\quad\left(k>0)\right.
\end{eqnarray}
But now $g\bar{g}$ has no time dependence and from (7)
$$ e^\zeta = \frac{kr}{sinh kr}$$
One obtains the famous selfdual BPS monopole with the magnetic
topological winding number
$$q = 1$$
Here we have, of course, the Euclidean version, $A_t$ replacing the
Higgs field. 

In (8) $g$ is precisely $Z_0$ the initial point of the iterative map
studied in App. Apply one iteration.  Then
\begin{equation}
g   = {{a_0 +e^{-k(r+it)}}\over{\bar a_0 + e^{k(r+it)}}} 
\end{equation}
{\it There is a spectacular change.} One has now an authentic periodic
solution characterized by {\it two} topological integers:

(1) The magnetic number remains unchanged since as $r \rightarrow\infty$ 
 there is no essential change in the configuration. One has still the
(monopole-like) number  
 $$ q = 1$$
 
(2) A second (instanton-like) topological integer $P_T$ is now given by
the total action $S_T$ over one period $(T= 2\pi k^{-1})$  divided by
$8\pi^2$. One defines
\begin{equation}
{8\pi^2} P_T = {4\pi}\int_0^T dt \int_0^\infty dr ({\partial_r}^2 + 
{\partial_t}^2)\omega
\end{equation}
where
$$\omega = {\frac{1}{2}}e^{2\zeta} + \ln
\Bigl(\frac{1 - g\bar g}{r}\Bigr)$$

\begin{equation}
({\partial_r}^2 + {\partial_t}^2)\omega = {\frac{1}{2}}({\partial_r}^2 +
{\partial_t}^2)(e^{2\zeta}-2\zeta) = {\frac{1}{2}}({\partial_r}^2 +
{\partial_t}^2)e^{2\zeta} + {\frac {1}{r^2}} (1-e^{2\zeta})
\end{equation}
For (9) one obtains
$$P_T = 2$$

The computation of the action ill be given below in a form particularly
suited to our purpose. But first let us introduce a more general form of
$g$.

In (8) set, assuming $k$ to be sufficiently large,
$$k = {\sum_{j=1}^{n}}k_j \qquad (k_j > 0)$$ 
At this stage the $k_j$'s are supposed to be mutually commensurate.
Thus ,in evident notations,
$$g = \displaystyle  \prod_{j=1}^{n} g_j^{(0)} = \displaystyle
 \prod_{j=1}^{n} e^{-k_j(r+it)}$$
Now apply one iteration to each factor giving
\begin{equation}
g = \displaystyle  \prod_{j=1}^{n} g_j^{(1)} = \displaystyle
 \prod_{j=1}^{n}\Biggl({{a_0^{(j)} +e^{-{k_j}(r+it)}}\over{\bar a_0^{(j)}
+ e^{{k_j}(r+it)}}}\Biggr) 
\end{equation}

In the notation of App
$$g_j^{(1)} =  Z_1^{(j)}$$
Each factor now has its own period
 \begin{equation}
T_j = 2\pi{k_j^{-1}}
\end{equation}
Since the $k_j$'s are mutually commensurate one can set
\begin{equation}
k_j = \hat{k} \frac{P_j}{Q_j}
\end{equation}
where $P_j , Q_j$ are integers without common factor. Thus there is an 
overall period 
\begin{equation}
T = \frac {2\pi}{\hat{k}} (\prod_{j} Q_j)
\end{equation}
We now compute the total action over $T$. Using (10) and Stoke's theorem
with the boundary indicated by the limits one obtains
\begin{equation}
S_T = 4\pi \int_{\delta}^R dr[\partial_t \omega ]_0^T + 4\pi \int_0^T
dt[\partial_r \omega ]_{\delta}^R  \qquad (\delta \rightarrow 0, R
\rightarrow \infty )
\end{equation}
The first integral vanishes due to periodicity. In the second one
non-zero contributions come only from the limit $r\rightarrow 0$. These
can be evaluated by using standard integrals [3,4]. {\it But the
very first step $(p = 0)$ of our study (App) of the circle map and
the small-r expansion furnishes the result directly.}      

From (12) and $(A.38)$ as $r\rightarrow 0$,
\begin{equation}
g\bar g = 1 + 2\bigl ( \sum_j \frac{d\psi_1^{(j)}}{dt}\bigr) r + 2\bigl (
\sum_j \frac{d\psi_1^{(j)}}{dt}\bigr)^2 r^2 + O(r^3)
\end{equation}
where
\begin{equation}
{e^{-i \psi _1^{(j)}}} = \Biggl({{a_0^{(j)} +e^{-i{k_j}t}}\over{\bar
a_0^{(j)} + e^{i{k_j}t}}}\Biggr) 
\end{equation}
From $(A.19)$ as
\begin{equation}
{k_j}t = \psi _0^{(j)} \rightarrow \psi_0^{(j)} + 2\pi
\end{equation}
$$\psi _1^{(j)} \rightarrow \psi_1^{(j)} + 4\pi$$
Inserting $\omega$ of (10) in (16) and using (17), the only non-zero
contributions are seen to come from
\begin{equation}
lim_{r \rightarrow 0}\Biggl (\partial_r \ln\Bigl(\frac{1 - g\bar
g}{r}\Bigr)\Biggr) = \sum_j \frac{d\psi_1^{(j)}}{dt}
\end{equation}
Hence, in terms of the previous definitions, one obtains quite simply
\begin{equation}
S_T = 4\pi\Bigl (4\pi \sum_{j=1}^n \frac {T}{T_j} \Bigl) = 8\pi T\bigl(
\sum_{j=1}^n k_j \bigr)
\end{equation}
$$= 8\pi{^2} 2  (\prod_{j} Q_j) \Bigl(\sum_{j=1}^n \frac
{P_j}{Q_j}\bigr)$$
Hence,
\begin{equation}
P_T = 2\Bigl ( \sum_{j=1}^n \frac {T}{T_j} \Bigl)
\end{equation}
Since each $\bigl ( \frac {T}{T_j} \bigl)$ is an integer $P_T$ is an 
even integer. Odd indices can be obtained through a simple modification
described below.

\subsection { Comments on the limiting values $a_0^{(j)}=0,\pm1$:}

In general the parameters $a$ , will be assumed to satisfy $0<\mid a\mid
<1$. The limits indicated are associated to monopoles and to standard
aperiodic instantons. The following points are to be noted.

(1) For
\begin{equation}
a_0^{(j)} = 0, \psi _1^{(j)} = 2k_{j}t
\end{equation}
This limit, consistent with (19), can be taken smoothly. For {\it all} 
$a_0^{(j)} = 0$ one has a static monopole. But having determined
beforehand each $T_j$ and $T$, if this limit is taken a posteriori one
obtains (formally) (22) in the simplest fashion.

(2) for
 \begin{equation}
a_0^{(j)} = 1, \psi _1^{(j)} = k_{j}t
\end{equation}
As compared to (23) there is a rescaling by $\frac {1}{2}$. ( The choice
$-1$ in (24) implies no essential change.) If say, only
$$a_0^{(1)} = 1$$
one obtains, instead of (22),
\begin{equation}
P_T =\frac {T}{T_1} + 2\Bigl ( \sum_{j=2}^n \frac {T}{T_j} \Bigl)
\end{equation}
Now $P_T$ can be odd. Consider, for example 
\begin{equation}
g=\Biggl({\frac{1}{2} (1+\epsilon) + e^{-k(r+it)}\over\frac{1}{2} (1+\epsilon) +
e^{k(r+it)\hphantom{-}}}\Biggr)
\prod_{j=1}^n\Biggl({a_j+e^{-k(r+it)}\over\bar a_j+e^{k(r+it)}}
\Biggr)
\end{equation}
 
For
\begin{equation}
\epsilon = -1, P_T = 2n + 2 ;\quad \epsilon = 1,  P_T = 2n + 1
\end{equation}

(3) For each $j$, taking the combined limit
\begin{equation}
k_j \rightarrow 0, a_0^{(j)} = -1 + k_{j}b_{j} \quad (b_j + \bar b_j > 0)
\end{equation}
one obtains from (12)
\begin{equation}
g =  \displaystyle
 \prod_{j=1}^{n}\Biggl({{b_j -(r+it)}\over{\bar b_j
+ (r+it)}}\Biggr) 
\end{equation}
This is Witten's multi-instanton solution [13] with centres on the time
axis and total action on $R^4$
\begin{equation}
S = 8\pi^2 (n-1)
\end{equation}
{\it The magnetic charge is lost in this limit.}

To sum up, $g$ as given by (12), combines instanton-like and
monopole-like aspects. It is a more general construction containing them
as limits. As an expression of this double role such periodic selfdual
configurations are characterized by the presence of {\it two}
topological integers: one instanton-like  $(P^T)$  and one monopole-like 
$(q)$. The combined role of these two in the index theorems will be
discussed later (Sec.6).
So far though $P_T$  can range through the entire spectrum of
integers,$q$ is restricted to unity. This restriction can be removed
(Sec.6).

Finally we just mention that concerning the rotation numbers defined in
App, the result (A.59) implies for (12), which has $n$ factors, a
staircase-like pattrn with $n$ steps. It was pointed in Sec.1 that
jumps in the rotation numbers can arise though the action density
remains smooth. We illustrate this explicitly for the simplest
non-trivial case.
Let
\begin{equation}  
g =  \Biggl({a + e^{-k(r+it)}\over\bar a + e^{k(r+it)}}\Biggr)
\end{equation}
Here the crucial value of $r$ corresponding to (A.59) is
\begin{equation}
e^{-kr} = a
\end{equation}
But now from (7),
\begin{equation}
e^{2\zeta} = (kr)^2\bigl ( (sinhkr)^{-2} - (1 - a^2)(coshkr + a
cost)^{-2}\bigr)
\end{equation}
Inserting (33) in (10) one obtains a smooth action density.

The results (A.39) and (A.40) can be used to study the time dependence of
the action density in a small sphere around $r = 0$. We defer the
discussion to be able to include higher iterations and quasiperiodicity.

\subsection {Higher iterations:}

For higher iterations one stays within the class of (quasi)periodic
instantons with unit magnetic charge. But various interesting properties
arise or are accentuated as the configuration becomes more complex. Some
crucial features are presented below.

   A generalization of the ansatz (12) is {\it automatically}
implemented. One more iteration on (9) gives (see (A.62))
\begin{equation}
g = Z_2 = Z_0^{-2}\Biggl (\frac {a_0 + Z_0}{\bar a_0 + Z_0^{-1}}\Biggr)
\Biggl (\frac {\mu_+ + Z_0}{\bar \mu_+ + Z_0^{-1}}\Biggr)\Biggl (\frac
{\mu_- + Z_0}{\bar \mu_- + Z_0^{-1}}\Biggr) 
\end{equation}
where
$$\mu_{\pm} = \frac {1}{2} \Bigl ( (a_0 + \bar{a_0}a_1) \pm \bigl ((a_0
+ \bar{a_0}a_1 \bigr)^2 - 4a_1 \bigr )^{\frac {1}{2}}\Bigr )$$
The factor $Z_0^{-2}$ (i,e, $e^{2k(r+it)}$) is an {\it increasing}
exponential in $r$, though $Z_2$ as a whole falls as $e^{-kr}$. This
aspect is generalized by setting (compare (12))  
\begin{equation}
g = e^{l(r+it)}\displaystyle\prod_{j=1}^{n}\Biggl({{a_0^{(j)}
+e^{-{k_j}(r+it)}}\over{\bar a_0^{(j)} + e^{{k_j}(r+it)}}}\Biggr) 
\end{equation}
with
$$l <  \bigl ( {\sum_{j=1}^{n}}k_j\bigr )$$
The total action, instead of (21), is now (assuming $l$ is so chosen that
the overall period is still $T$)
\begin{equation}
S_T = 4\pi T\bigl(2\sum_{j=1}^n k_j -l \bigr)
\end{equation}
Thus one has a lower action with the same number of parameters $a$. In
(34) the factor $Z_0^{-2}$ lowers $S_T$ (and $P_T$) to give 
\begin{equation}
(8\pi^2)^{-1} S_T = P_T = 2(3-1) =4
\end{equation}
 \subsection {Effect on rotation numbers:}
The rotation numbers are defined and discussed in App ((A.58) to (A.65)).
In the course of time evolution of the gauge field generated by (34),$g$
can vanish for 
\begin{equation}
e^{-kr} = \mid a_0 \mid,  \mid \mu_+ \mid,  \mid \mu_- \mid
\end{equation}
though $\mid \mu_+ \mid$ and  $\mid \mu_- \mid$ may coincide to $\mid
a_1 \mid ^\frac{1}{2}$.
Thus though $S_T$ is doubled, the number of zeros are here (generically)
tripled. 

Let $r_1 < r_2 < r_3$ be the three distinct roots of (38) in $r$ and let
$\Omega_2$ denote the rotation number after the second iteration. Then
(A.59) is generalized as follows.
\begin{equation}
\frac {\Omega_2}{\Omega_0} = 4,\frac {7}{2},3,\frac {5}{2},2,\frac
{3}{2},1 
\end{equation}
respectively for
$$r <r_1, r = r_1, r_1 <r <r_2, r = r_2, r_2 <r <r_3, r = r_3, r >
r_3.$$

Cumulative effects of higher iterations will increase the number of
steps. One can factorize $Z_{p+2}$ in terms of $Z_p$ analogously to
(34). But rather than repeating such steps our aim is to show how the
results of App can lead directly to remarkable properties. An example,
in a different direction, follows.
\subsection {Action density near the spatial origin:}

Let us explore the role of the iterations in the action density in a
small sphere about the origin where the intricate interplay of
(quasi)periodic pulsations are least affected by exponential damping
with increasing $r$.
We generalize (12) by setting
 \begin{equation}
g = \prod_{j=1}^n Z_{p_j}^{(j)}
\end{equation}
where $Z_{p_j}^{(j)}$ is $e^{-k_j{(r+it)}}$ iterated $p_j$ times. For
each $p_j=1$ one obtains (12).
The doubling after each iteration of the contribution of each factor to
the total action has to be taken into account. A simple particular
case is given by (37). The general result is evident on noting how
(A.19) generalizes (19) and hence the derivation of (21). For (40) one
obtains, instead of (21), 
\begin{equation}
S_T =  8\pi T\bigl(\sum_{j=1}^n 2^{(p_j-1)} k_j \bigr)
\end{equation}
As the action increases, the "weight" of that particular field
 configuration in path integrals diminishes. Hence the interest of an
extra factor, as in (35), bringing down the action as far as possible
for a given period, as in (36).
Having obtained the total action let us now take a close look at a small
sphere around $r=0$. Using the results obtained in (A.38),(A.39) and
(A.40) one obtains after simplifications for (40)
\begin{eqnarray}
e^{2\zeta} = \frac{r^2}{(1 - g\bar{g})^2} \Bigl( ({\partial_r}^2 + 
{\partial_t}^2)(g\bar{g})\Bigr) = 1 - \frac {2}{3} \Gamma_t r^2 + O(r^3)
\end{eqnarray}
where
\begin{eqnarray}
 \Gamma_t = &\frac {1}{2} C^2 + 
({\ddot{C}}/{C})
  -  \frac {3}{2} \bigl ( {\dot{C}}/{C}\bigr)^2 \\
  C = &\sum_{j=1}^n {\dot {\psi}}_{p_j}^{(j)} \equiv \dot {\Psi} 
\end{eqnarray}
  Here the dots denote time derivatives and $\psi_{p_j}^{(j)}$ is the
phase  of the j-th factor (for $r=0$) in (40). This leads, for
$\omega$  given by (10) 
\begin{equation} 
({\partial_r}^2 + {\partial_t}^2)\omega = \frac {4}{3} \Gamma_t r^2 +
O(r^3)
\end{equation}
The total action over $R^3$ and a period $T$ is given by (41). Let $S(r)$
denote the action, at any instant $t$, integrated over a small sphere
of radius $r$ about the origin. Then  
\begin{equation}
S(r) = 4\pi \int _0^r dr ({\partial_r}^2 + {\partial_t}^2)\omega 
= \bigl( \frac {4}{3} \Gamma_t^2 \bigl)V_r + \ldots
\end{equation}
where (the dots indicating higher powers of $r$) 
$$V_r = \frac {4}{3} \pi r^3$$

The leading term, logically, is proportional to the volume of the
sphere.(This is rendered possible by the zero coefficient of the term
linear in $r$ in (42).)

In (43) $\Gamma _t$ has an interesting structure. There is a "kinetic"
term,
$$\frac {1}{2} C^2 = \frac {1}{2}\dot {\Psi}^2 $$
and the {\it Schwarzian derivative} of $\Psi$. For the simplest case
(one factor with $p=1$ in (40)) using (A.17), with
$$a=\mid a_0\mid ,\chi = \psi_0 -\alpha_0 , \psi_0 = kt$$
$$C=\dot {\Psi}= 2k\frac {1+a cos\chi}{1+a^2+2a cos\chi} \equiv 2k\frac 
{X}{Y}$$
 Hence
\begin{equation}
\frac {\ddot{C}}{C} - \frac {3}{2}\Biggl( \frac {\dot{C}}{C}\Biggr)^2 = 
k^2 a (1 - a^2) \bigl ( cos\chi + a (\sin\chi)^2\bigl (\frac {3}{2}
X^{-1} + Y^{-1}\bigr) \bigr) (XY)^{-1}
\end{equation}
This changes sign at points determined by the choice of $a_0$. But for
all $a_0$, it is positive (negetive) for $\chi =0 (\pi)$ respectively. 
For the general case (40) the very complex structure implied by (43)
determines the time dependence.

\section{Quasiperiodicity:}

So far we have been considering strict periodicity, having just
mentioned that the component periods $T_j$ need not necessarily be
commensurate. We now take a closer look at this possibility.

As an example let $(k_2, k_3, \ldots , k_n)$ in (12) be all mutually
commensurate as in (14) but {\it not} with $k_1$. ( This relatively
simple case will suffice to exhibit some remarkable features. It is
possible to consider several incommensurate periods. One can also start
with $Z_0$ given by (3) rather than by (2).)

Let $\hat{T}$ be the common overall period for $(T_2, T_3,\ldots, T_n)$,
but {\it not} for $T_1$. One can take successive rational approximations
of
$\bigl (\frac {\hat{T}}{T_1}\bigr)$
\begin{equation}
 \Bigl (\frac {\hat{T}}{T_1}\Bigr)_{(appr)} = \frac {N_1}{N_2}
\end{equation}
$N_1, N_2$ being integers without common factors. Then
$$T = N_2 \hat{T}$$
is the overall period at that approximation. As the approximation is
improved $N_1, N_2, T$ all increase without limit. ( Since there is no
exact period, improving the approximation one tends to cover the entire
time axis.)
One can keep in mind the famous example of successive rational
approximations of the "golden mean"
$$G = \frac {1}{2} \bigl( \sqrt{5} -1 \bigr)$$
using the Fibonacci sequence. But our considerations will not be limited
to any such particular case. For the $l$-th approximation, in evident
notations, the action over  the period $T^{(l)}$ is

\begin{equation}
S_l = 16{\pi}^2 \Biggl ( \frac {T^{(l)}}{T_1} + \sum_{j=2}^n \frac
{T^{(l)}}{T_j} \Biggr )
\end{equation} 

As $l \rightarrow \infty$ so do $T^{(l)}$ and $S_l$. One can now consider
the limiting form of the action per unit time or the "normalized" action
\begin{equation} 
S_N = \Biggl(\frac {S_l}{T^{(l)}} \Biggr)_{l\rightarrow \infty}
\end{equation}This gives a {\it continuous index}, a positive real
number, not an integer. The possible mathematical significance of $S_N$
has been discussed elsewhere [ 4,7].
The magnetic index $q$ does {\it not} vary in the successive stages
described above. As one approaches the asymptotic $S_2$ in $R^3$ the
exponential damping of time dependences give the same static
configuration at each stage, namely that of the monopole. Hence our
{\it quasiperiodic instantons are characterized, not by two topological
integers, but by one positive real number $(S_N)$ and one integer
$(q)$.} For the class of solutions under consideration $q = 1$. More
general possibilities are indicated in Sec.6.

So far we have been looking at quasiperiodic solutions from the point
of view of successive approximations. Let us now look at some {\it
exact} consequences of the postulated incommensurability.

In the case considered before, consider the Poincar\'e sections of the
factor $g_1$ ( corresponding to the period $T_1$ ) for
$$t = t_0 + nT \quad (n = 0,1,2, \ldots )$$
One obtains
\begin{equation}
g_1 (n) = e^{-i2\alpha_1}{{\mid a_1\mid + e^{-k_1(r +
i(t_0 + nT)) + i\alpha_1}}\over {\mid a_1 \mid + e^{\ k_1(r+i(t_0 + nT))
-i\alpha _1}}}
\end{equation}
We simplify notations by setting
$$a = \mid a_1\mid = a_1 e^{i\alpha_1}, k_1 = k, k\omega = kt_0
-\alpha_1$$
$$X^{(n)} =  e^{i2\alpha_1}g_1 (n) = \frac {a + F_n}{a + F_n^{-1}}$$
where
$$F_n = e^{-k(r+i\omega+inT)} = e^{-k(r+i\omega) -i2\pi n\delta},\
\delta \neq \frac {N_1}{N_2} \quad (N_1,N_2\  integers)$$
{\it The irrationality of $\delta $ is the basic quasiperiodicity
postulate.}
Now 
\begin{equation}
X^{(n+p)} =   \frac {a + F_{n+p}}{a + F_{n+p}^{-1}} = \frac {a +
F_n e^{-i2\pi p\delta}}{a + F_n^{-1}e^{i2\pi p\delta}}
\end{equation} 

We show that if the parameters $(r, t_0, k, \alpha, n, p, \delta)$ are
fine-tuned $X$ may come back {\it just once} to an initial value but no
more ({\it not even twice}). From (52)
\begin{equation}
F_n^2 + e^{i2\pi p\delta} a (1 - X^{(n+p)}) F_n - e^{i4\pi
p\delta}X^{(n+p)} =0 
\end{equation}
Eliminating $F_n$ from the equations for three {\it distinct} integer
values of $p$ say,
$$p = p_1, p_2, p_3$$
one of which may be zero, one has as a necessary constraint the
vanishing determinant
\begin{equation}
{\rm det}\displaystyle \left\vert\matrix{1 & a( 1 - X^{n+p_1})e^{i2\pi
p_1
\delta} & - X^{n+p_1}e^{i4\pi p_1 \delta} \cr 1 & a( 1 -
X^{n+p_2})e^{i2\pi p_2
\delta} & - X^{n+p_2}e^{i4\pi p_2 \delta} \cr 1 & a( 1 -
X^{n+p_3})e^{i2\pi p_3 \delta} & - X^{n+p_3}e^{i4\pi p_3
\delta}}\right\vert = 0
\end{equation}
For, say,
\begin{equation}
X^{(n+p_1)} = X^{(n+p_2)} =   X^{(n+p_3)} = X
\end{equation}
this reduces to
\begin{equation}
X(1 -X) e^{i(p_1 + p_2 +p_3)2\pi \delta} \Bigl (\bigl (1 - e^{i(p_1 -
p_2 )2\pi \delta}\bigr ) \bigl (1 - e^{i(p_2 -
p_3 )2\pi \delta}\bigr ) \bigl (1 - e^{i(p_3 -
p_1 )2\pi \delta}\bigr ) \Bigr ) = 0
\end{equation}
The coefficient of $X(1 -X)$ cannot vanish due to our quasiperiodicity
postulate. Neither can $X^{(n+p)}$ be zero or unity for three distinct
values of $p$. That would imply, for example for $X = 1$,
$$F_n^2 = e^{i4\pi p_j \delta}$$ 
for three different values of $p_j$ and so on. Hence the constraint (55)
cannot be satisfied. So $X^{(n)}$ cannot come back twice {\it exactly} to
a previous value. There is, however, no restriction concerning repeated
very close approaches.

Let us now look at the conditions necessary for a {\it single} return,
namely
$$X^{(n+p)} = X^{(n)}$$
From (52) this is seen to imply
\begin{equation}
\bigl (F_n e^{-i\pi p\delta}\bigr)^2 + 2\lambda \bigl (F_n e^{-i\pi
p\delta}\bigr) +1 = 0
\end{equation}
where
$$\lambda = a^{-1} cos\pi p\delta$$
Hence 
\begin{equation}
F_n = e^{i\pi p\delta}\bigl ( -\lambda \pm \sqrt{\lambda^2 - 1}\bigr)
\end{equation}

   $${\rm Case (1) :} \quad (\mid \lambda \mid \leq 1)$$

Let $\lambda = cos \eta \quad (\eta {\rm real})$. Then
\begin{equation}
F_n = e^{-k(r+i\omega ) -i2\pi n\delta} = - e^{i(\pi p\delta \ \pm\ 
\eta)}
\end{equation}
This implies $r = 0$ and 
\begin{equation}
e^{-i(kt_0 - \alpha_1\  +\  \pi (2n\  +\  p)\delta \ \pm\  \eta )} = -1 
\end{equation}
(The $\pm$ sign implies one or the other. They do not hold
simultaneously.)

 $${\rm Case (2) :} \quad (\mid \lambda \mid > 1)$$

Let $\lambda = coth\zeta $ ($\zeta$  real). Then, retaining the
solution leading to a real value of $r$,
\begin{equation}
F_n = - \bigl (tanh(\zeta /2)\bigr) e^{i\pi p\delta}
\end{equation}
and
\begin{equation}
e^{-kr} = \mid tanh(\zeta /2) \mid ,\quad e^{-i(kt_0 -\alpha_1\  +\  \pi
(2n\  +\  p)\delta) } = \pm 1
\end{equation}

It should be noted that $X^{(n+p)} = X^{(n)}$ does {\ not} imply a
periodic situation. In the solutions above $n$ is not arbitrary for a
given $p$. One can indeed show that given the foregoing situation
$$ X^{(n+2p)} \neq X^{(n+p)} = X^{(n)}$$
consistently with the impossibility of (55). In this context the
ambiguity of sign of the inverse map (expressing $F_n$ in terms of $X_n$)
plays an interesting role. But we will not present such details here.

Let us note one more point.{\it  The action density and the rotation
numbers now involve mutually incommensurable numbers.} It would be
interesting to study in detail the time evolution of the coefficient (43)
\begin{eqnarray}
 \Gamma_t = &\frac {1}{2} C^2 + 
({\ddot{C}}/{C})
  -  \frac {3}{2} \bigl ( {\dot{C}}/{C}\bigr)^2 \\
  C = &\sum_{j=1}^n {\dot {\psi}}_{p_j}^{(j)} \equiv \dot {\Psi} 
\end{eqnarray}
with three or more incommensurable periods (namely $k_j$'s at the origin
of the $\psi_j$'s) varying the parameters $a$.
Our purpose in leaving the period $T$ and $\Omega_0$ in (A.58) is now
evident. In general one will have, to start with, incommensurable
$\Omega_0^{(j)}$'s corresponding to different factors of $g$. For the
simple case considered above (only $k_1$ incommensurable with the rest
and with $F_n$ corresponding to a section of $Z_0^{(1)}$) one has 
$$\Omega_0^{(1)} = 2\pi \delta$$
$\delta$ being irrational. The cumulative effects of several irrational
numbers have to be taken into account in a more general case.

The different roles of rational and irrational rotation numbers are
well-known [9,10]. Here the rotations of the phases are considered in
the context of annular maps (App). The "average" doubling of the phase
(sufficiently near the origin) for each iteration is also specific to
our case. Though we have demonstrated the existence of striking
discontinuities, the role of the rotation numbers remains to be explored.

\section{Index theorems and spinor solutions:}

Again we start with the periodic case, defering a discussion of
quasiperiodicity.
For periodic gauge fields the base manifold is $R^3\times S^1$ rather
than $R^4$ or $S^4$. The index theorems have to take into account the
boundary effects induced by $S^1$. This has been discussed elsewhere
[4,6,7]. Here we just mention that the number of zero modes of spinors
in periodic backgrounds characterized by two topological integers $(P_T,
q)$

\        $=\ P_T\ -\ q$ for periodic spinors of isospin $\frac
{1}{2}$\  (see [2,4])

\        $=\ 4P_T\ -\ 2q$ for periodic spinors of isospin $1$ (see
[7])

\        $=\ P_T$ for antiperiodic spinors of isospin
$\frac{1}{2}$ (see [6])

(The background is, of course, still perioˆdic for antiperiodic
spinors.)  
\ 
We construct below the spinor solutions in a more general and
systematic fashion than in the previous papers. This is facilitated
by introducing a gauge transformation leading from the Witten-type
gauge introduced in (5) and (6) to a (quasi)periodic generalization
of 't Hooft or Jackiw -Nohl -Rebbi (JNR) type solutions [2,4,6,7].
Two historical references for the standard (aperiodic) solutions are
[13,14].    
\subsection {Gauge transformation :}
We start with $A_\mu$ given by (5) and (6) and gauge transform by
$$G(r,t) = e^{-id(r,t)\sigma_r /2}$$
where
\begin{equation}
e^{id} = \Biggl(\frac {1 -g}{1 -\bar g}\Biggr)\Biggl (\frac {(\partial_r
-i\partial_t)g}{(\partial_r
+i\partial_t)\bar g}\Biggr)^{\frac{1}{2}}
\end{equation}
One obtains
$$A_\mu^\prime = G^{-1}A_{\mu} G - i G^{-1}\partial_{\mu} G =
\sigma_{\mu \nu} \partial_{\nu} ln \Sigma$$
where $\sigma_{10} =\sigma_{23} =\sigma_{1} \  {\rm (cyclic)}$, and
\begin{equation} 
\Sigma = \frac {1}{2r} \Biggl(\frac {1 +g}{1 - g} + \frac
{1 +\bar g}{1 -\bar g}\Biggr)\  =\  \frac{1}{r}\ \frac {(1 - g\bar g)}{(1
- g)(1 - \bar g)}
\end{equation}
satisfying ,since  $(\partial_r + i\partial_t)g = 0 = (\partial_r -
i\partial_t)\bar g$,
\begin{equation}
\displaystyle {\Box}  \Sigma =\Bigl (\partial_t^2 + \partial_r^2 + \frac
{2}{r}
\partial_r\Bigr)\Sigma = 0
\end{equation}
Thus we have obtained the famous ansatz leading to 't Hooft or JNR
solutions. But here it is being implemented in the context of
magnetically charged (quasi)periodic solutions rather than that of
standard aperiodic instantons.

The poles of $\Sigma$ play an essential role in the construction of
spinor solutions. {\it The passage from $g$ to $\Sigma$ maintains contact
with the iterative map implemented for $g$}. But this raises the problem
of displaying in an additive form the poles of $\Sigma$. The roots of $g
= 1$ are not ,in general, explicitly available.
Genrally there are $2n$ roots for (12) ,all for $r = 0$. The particular
case, with $(0 <a <1)$ and
$$g = \Biggl (\frac {a + e^{-(r+it)}}{a + e^{(r+it)}}\Biggr )^n$$
is fully treated in [7].
We display some typical features of the passage from $g$ to $\Sigma$
through simple examples, motivating the generalization to follow.
\smallskip

\underline{(1)} \  For \ $g=e^{-k(r it)}$

$$\frac{1+g}{1-g}=coth\frac{1}{2} k(r+it)$$
\begin{equation}
\Sigma = \frac{1}{2r}\Bigl(coth\frac{1}{2} k(r+it)  + coth\frac{1}{2}
k(r-it)\Bigr) 
\end{equation}
$$=\sum_{l=-\infty}^{\infty} \frac {1}{k^2} \frac {1}{r^2+(t-k^{-1}2\pi
l)^2}$$
This is the periodic form of the BPS monopole, gauge equivalent with the
still better-known static form. The action over one period $(T=2\pi
/k)$can be considered, quite formally, to be $8\pi^2$ (or $P_T=1)$. (See
the comments following (8) and (23).)

\smallskip

\underline{(2)}\  For, with $(0 <a <1)$,
$$g = \Biggl (\frac {a + e^{-k(r+it)}}{a + e^{k(r+it)}}\Biggr )$$
$$P_T=2,$$
and 
\begin{equation}
\frac{1+g}{1-g}=\frac{1+a}{2} coth\frac{1}{2} k(r+it) +\frac{1-a}{2}
coth\frac{1}{2} k(r+it - i\pi k^{-1})
\end{equation}

\smallskip

\underline{(3)}\  For   
$$g = e^{-k(r+it)} \Biggl (\frac {a + e^{-k(r+it)}}{a +
e^{k(r+it)}}\Biggr )$$
$$P_T=3 \quad     {\rm (The\  simplest\  case\  of\  (26)\  and \
(27))}$$ and
\begin{equation}
\frac{1+g}{1-g}=\lambda_1^2 coth\frac{1}{2} k(r+it) +\lambda_2^2
coth\frac{1}{2} k(r+it - ic k^{-1}) + \lambda_3^2
coth\frac{1}{2} k(r+it + ic k^{-1})
\end{equation}
where
$$cosc=-\frac{1+a}{2} ;\ \lambda_1^2 = \frac{1+a}{3+a},\  \lambda_2^2 = 
\lambda_3^2 = \frac{1}{3+a} $$

For our normalization a general consequence is
$$\sum_i \lambda_i^2=1$$

The number of $coth$ terms in $\frac{1+g}{1-g}$ gives $P_T$ if each one
has the same period, say $T$ and {\it distinct singularities}. A term
with period $(T/m)$ contributes $m$ units to $P_T$.
\smallskip

\underline{(4)} \quad { Once the roots of $g=1$ are obtained the
residues at the poles give the $\lambda$'s. But this is not strictly
necessary for constructing the spinor solutions.}

Consider a case with two different periods ($2\pi k^{-1},\pi k^{-1}$),
namely, $(0<a<1)$ and

$$g = \Biggl (\frac {a + e^{-k(r+it)}}{a + e^{k(r+it)}}\Biggr )\Biggl
(\frac {a + e^{-2k(r+it)}}{a + e^{2k(r+it)}}\Biggr )$$
Now 
\begin{equation}
P_T=2+2.2=6,\quad q=1
\end{equation}
The zeros of $g$ (and hence the jumps in rotation numbers) correspond to
$$e^{-kr}=a,\sqrt a$$ 
It is sufficient to note that $g=1$ for
$$r=0;\quad e^{it}=\pm 1,\Biggl(-\frac{1}{2} \pm i\frac{\sqrt
3}{2}\Biggr),
\Biggl(\frac{1-a}{2} \pm i\sqrt{ 1- \Bigl(\frac{1-a}{2}\Bigr)^2}\Biggr)$$
One can then construct $(P_T-q)=5$ periodic and $P_T=6$ antiperiodic
spinor solutions. The explicit solutions will follow.

More generally, using real parameters $a_p$ at each iteration one
obtains some simplifications. Thus, for example, iterating $g$ as a
whole (with $0<a_p<1$)
\begin{equation}
g_{p+1}=\frac{a_p+g_p}{a_p+g_p^{-1}}
\end{equation}
and
\begin{equation}
 \frac{1+g_{p+1}}{1-g_{p+1}}=-\Biggl(\frac{1+a_p}{2}\Bigl(
\frac{1+g_p}{1-g_p}\Bigr)+\frac{1-a_p}{2}\Bigl(
\frac{1-g_p}{1+g_p}\Bigr)\Biggr)
\end{equation}

Hence\quad $g_{p+1}=1$ \quad for $g_p=\pm1$, consistently with the
doubling of the periodic action.
For real $a_0,a_1$ and
$$g_0=e^{-k(r+it)}$$
$$g_2=1\quad {\rm for}\quad r=0;\quad sin\frac{kt}{2}=0,
cos\frac{kt}{2}=0,\pm \sqrt{\frac{1-a_0}{2}}$$
\subsection{Generalization of $Z_0$ and $g$:}
Having noted the relation between the structures of $g$ and $\Sigma$ one
can now invert the procedure. Start with
\begin{eqnarray}
\Sigma &= &\frac{1}{2r}\sum_{l=1}^n \lambda_l^2 \Bigl(coth\frac{1}{2} k_l
(r+it -ic_l)  + coth\frac{1}{2} k_l (r-it +ic_l)\Bigr) \\
&= &\frac{1}{2r}\Bigl(\frac{1+g}{1-g}+\frac{1+\bar g}{1-\bar g}\Bigr)
\end{eqnarray}
where convenient conventions are
$$\sum_{l=1}^n \lambda_l^2 = 1;\quad k_n > k_{n-1} >\ldots >k_2 > k_1 >
0$$
Now
\begin{equation}
g=\frac {\sum_{l=1}^n \lambda_l^2 \Bigl(e^{k_l (r+it -ic_l)} -
1 \Bigr)^{-1}}{\sum_{l=1}^n \lambda_l^2 \Bigl(1-e^{-k_l (r+it -ic_l)}
 \Bigr)^{-1}}
\end{equation}
Thus we have motivated the parametrization (3) of $Z_0$ (here $g$). One
can verify that (76) satisfies all the constraints listed below (7). In
particular, $g$ is a phase for $r=0$ and drops as $e^{-k_1 r}$ as $r$
becomes large. {\it The major interest is that now, even in presence of
an arbitrary number of different periods (different $k$'s) $g$ is
directly adapted to the construction of spinor solutions to follow.}
The $\Sigma$ of (74) can easily be generalized to break spherical
symmetry. But we defer such considerations to Sec.6.

To give a better idea of the relation of (76) to the previous
factorized form we cosider again some simple examples. Different
subclasses of the previous form will be found already in the simplest
examples of (76).

           We set $n=2;\quad c_1=c_2=0$ and we use below often
$\lambda_1^2 + \lambda_2^2 = 1$.

\smallskip
 
\underline{(1)}\quad  For $k_1=k_2=k$
$$g=e^{-k(r+it)}$$
More generally, for all $k$'s and all $c$'s equal one has the monopole
solution.

\smallskip
 
\underline{(2)}\quad  For $k_1=k ,k_2=2k$
$$g=\frac {\lambda_1^2 + e^{-k(r+it)}}{\lambda_1^2 + e^{k(r+it)}}$$

\smallskip

\underline{(3)}\quad  For $k_1=k ,k_2=3k$
$$g=e^{k(r+it)}\Biggl(\frac {a+e^{-k(r+it)}}{\bar a + e^{k(r+it)}}\Biggr)
\Biggl(\frac {\bar a + e^{-k(r+it)}}{ a + e^{k(r+it)}}\Biggr)$$

where $a=\frac{1}{2}\Bigl(\lambda_1^2 + i\lambda_1
\sqrt{4-\lambda_1^2} \Bigr) \quad \bigl(a\bar a = \lambda_1^2 <1 \bigr)$

\smallskip

\underline{(4)}\quad  For $k_1=2k ,k_2=3k$
$$g=\Biggl(\frac {a_1+e^{-k(r+it)}}{\bar a_1 + e^{k(r+it)}}\Biggr)
\Biggl(\frac { a_2 + e^{-k(r+it)}}{ \bar a_2 + e^{k(r+it)}}\Biggr)$$
where respectively for
$$1>\lambda_1^2>\frac{1}{4}, \quad a_1=\bar
a_2=\frac{1}{2}\Bigl(1+i\sqrt{4\lambda_1^2 - 1}\Bigr) ;$$

$$\frac{1}{4}  >\lambda_1^2 >0,\quad a_1=\bar
a_1=\frac{1}{2}\Bigl(1+\sqrt{1 -4\lambda_1^2 }\Bigr) ,$$ 
$$a_2=\bar a_2=\frac{1}{2}\Bigl(1-\sqrt{1 -4\lambda_1^2 }\Bigr);$$
$$\lambda_1^2 =\frac{1}{4},\quad a_1=a_2=\frac{1}{2}.$$

{\it For special values of the $k$'s and the $c$'s in (76) some poles can
coincide, diminishing $P_T$ accordingly.} The preceding examples, apart
from leading to simple factorized expressions for $g$ also illustrate
this point. One degeneracy (common pole for $r=0$, $sinkt=0$) diminishes
$P_T$ by $1$ in each case. This is consistently incorporated in the
factorized forms of $g$. Thus for $k_1=2k ,k_2=3k$ one obtains $P_T=4$
instead of 5. We will assume, in general,that the choice of parameters
in $\Sigma$ imply distinct poles.

Spinor solutions for zero mass were constructed [2] directly in Witten
type gauge. But since the counting of the number of zero modes (and hence
comparison with the index theorems) is more transparent in the 't Hooft
or JNR type gauge, we start by generalizing (74) as follws. Let
$${A_\mu} = \sigma_{\mu \nu} \partial_{\nu} ln \Sigma$$
where, with $\epsilon=\pm1$,

 $$\Sigma = \frac{1}{2}(1+\epsilon)+\frac{1}{2r}\sum_{j=1}^n
\lambda_j^2
\Bigl(coth\frac{1}{2} k_j (r+it -ic_j)  + coth\frac{1}{2} k_j (r-it
+ic_j)\Bigr) $$ 
\begin{equation}
 = \frac{1}{2}(1+\epsilon)+\sum_{j=1}^n\sum_{l=-\infty}^{\infty} \frac
{\lambda_j^2}{k_j^2}\frac {1}{r^2+(t-k_j^{-1}2\pi l)^2}
\end{equation}
The role of $\epsilon$ is interesting concerning the index theorems. For
$\epsilon=1$ one has a 't Hooft-like form. It can be shown that on the
asymptotic sphere in $R^3$ one has now a dipole like configuration
rather than a monopole.( See the discussion in [6] and the sources cited
there.) Hence for
$$\epsilon = 1,\quad q=0$$
while for
$$\epsilon = -1,\quad q=1.$$

However, for {\it both} values $(\epsilon = \pm 1)$, the periodic action
is the same. For an overall period $T$ and integers $n_j$ such that
$$T_j=2\pi k_j^{-1},\quad T=2\pi k^{-1},\quad k_j=n_j k$$
\begin{equation}
S_T=8\pi^2\sum_j n_j
\end{equation}
This can be seen as a limiting case of the familiar ('t Hooft and JNR)
solutions as follows. Start by retaining in (77) the poles covering an
interval $NT$. Define, in evident notations,
$$S_T=\bigl(N^{-1}S_{NT}\bigr)_{N\rightarrow \infty}$$
The value of $S_{NT}$ is given by the very well-known results for the 't
Hooft and JNR solutions leading to
$$S_T= \frac{8\pi^2}{N}\bigl(N\sum_j n_j - (1-\epsilon)/2
\bigr)_{N\rightarrow \infty}$$
$$=8\pi^2 \sum_{j=1}^n n_j$$
{\it Thus for the periodic case, in contrast with the aperiodic one, the
change from $\epsilon =-1$ to $\epsilon =1$ does not increase $P_T$ by
$1$ but diminishes $q$ by $1$.}
For $n=1$ one has quite a special case.Then for $\epsilon=-1$ one has a
static monopole gauge transformed to a periodic form,whereas for
$\epsilon=1$ the periodicity is authentic but $q=0$.

\subsection {Spinor solutions:}
Having thoroughly prepared the ground, we at last introduce the spinors.
Here we consider  [2,6] only isospin
$$I=\frac{1}{2}$$
( See [7] for $I=1$.) We consider only zero mass spinors.

For our conventions only upper (positive) helicity spinors have
normalizable solution. Separating this helicity $( \Psi_U)$ the Dirac
equation reduces to
$$\bar \alpha_{\mu} \bigl(i\partial_{\mu} - A_{\mu} \bigr) \Psi_U = 0$$
where
$$\bar
\alpha_{\mu}=\bigl(\tau_0,i\vec{\tau}\bigr),\quad \alpha_{\mu}
=\bigl(\tau_0,-i\vec{\tau}\bigr)$$
(We denote space-time Pauli matrices by $\tau_\mu$ and those in the
isospace by $\sigma_\mu$.)
The isospin components are separated as
$$\Psi_U= \displaystyle \left\vert\matrix{\Psi_U^{(+)} \cr
\Psi_U^{(-)}}\right\vert, \quad I_3\Psi_U^{(\pm)}=\pm
\frac{1}{2}\Psi_U^{(\pm)}$$
Set
$$\Psi_U^{(\pm)}=\Sigma^{\frac{1}{2}}\displaystyle
\left\vert\matrix{a_{\pm} \cr b_{\pm}}\right\vert$$
and define
$$u=\frac{1}{2}(x_3 + ix_0),\quad \bar u=\frac{1}{2}(x_3 - ix_0)$$
$$v=\frac{1}{2}(x_1 + ix_2), \quad \bar v=\frac{1}{2}(x_1 - ix_2)$$
Formally, the solutions are given (see [6] and the sources cited there)
by
$$a_+=\partial_v \bigl(\Sigma^{-1} H\bigr),\quad b_+=-\partial_u
\bigl(\Sigma^{-1} H\bigr)$$
$$a_-=-\partial_{\bar u} \bigl(\Sigma^{-1}
H\bigr),\quad b_-=-\partial_{\bar v}
\bigl(\Sigma^{-1} H\bigr)$$
where
$$\Box H= (\partial_u \partial_{\bar u} + \partial_v \partial_{\bar
v}) H=0$$
{\it Normalizable} solutions are obtained by matching the poles of $H$
with the zeros of $\Sigma^{-1}$.From (77) it is evident that there are
an infinite number of such possible choices of $H$. A finite number is
obtained by imposing suitable boundary conditions relating $\Psi_U (t)$
and $\Psi_U (t+T)$, $T= 2\pi K^{-1}$ being the period.

For {\it periodic} spinors satisfying
$$\Psi_U (t) = \Psi_U (t+T)$$
set
\begin{equation}
H_{m_j}=\frac{1}{2r}\Bigl(coth\frac{1}{2} K\bigl(r+i(t-c_j)-i2\pi
k_j^{-1} m_j\bigr)+coth\frac{1}{2} K\bigl(r-i(t-c_j)+i2\pi
k_j^{-1} m_j\bigr)\Bigr)
\end{equation}
with $T_j= 2\pi k_j^{-1}$ and
$$m_j=1,2,\ldots ,n_j;\quad k_j=n_j K;\quad (j=1,2,\ldots ,n)$$
These provide all the normalizable  zero modes. The total number is
\begin{equation}
\Biggl(\sum_{j=1}^n n_j\Biggr)
\end{equation}
But they are not all necessarily independent since
\begin{equation}
\frac{1}{n_j} \sum_{m_j=1}^{n_j} coth\frac{1}{2}
K\bigl(r\pm i(t-c_j-2\pi k_j^{-1} m_j)\bigr) = coth\frac{1}{2}
k_j\bigl(r\pm i(t-c_j)\bigr)
\end{equation}
(This is probably most easily derived via the logarithmic derivative of
the well-known product formula for $sinhnx$.)
Hence
\begin{equation}
\sum_j n_j^{-1} \lambda_j^2\Biggl(\sum_{m_j} H_{m_j}\Biggr)=\Sigma
-\frac{1}{2} (1+\epsilon)
\end{equation}
Thus for $\epsilon = -1$, in an evident notation,
 \begin{equation}
\sum_j n_j^{-1} \lambda_j^2\Biggl(\sum_{m_j} \Psi_U^{(m_j)}\Biggr)=0
\end{equation}
{\it There is no such constraint for $\epsilon=1$.}
Moreover, for {\it both} cases $(\epsilon=\pm 1)$, as explained before
$$P_T=\Biggl(\sum_{j=1}^n n_j\Biggr)$$
Thus in both cases, consistently with the index theorems stated at the
beginning of this section,the number of zero modes
$$=P_T-q.$$

Here $q$ is limited to zero and one. But, on the other hand, we have
done much more than counting the number of possible solutions. They have
been obtained explicitly. One can now study, to take only one example,
the time evolution of such spinor densities.

For {\it antiperiodic} spinors (in periodic backgrounds)
$$\Psi_U (t) = -\Psi_U (t+T)$$
and the correct choice [6] for $H$ turns out to be
\begin{equation}
H_{m_j}=\frac{1}{2r}\Bigl(cosech\frac{1}{2} K\bigl(r+i(t-c_j)-i2\pi
k_j^{-1} m_j\bigr)+cosech\frac{1}{2} K\bigl(r-i(t-c_j)+i2\pi
k_j^{-1} m_j\bigr)\Bigr)
\end{equation}
The linear constraint (83) is now absent even for $q=1$ ($\epsilon= -1$).
{\it There is no "magnetic defect"},no subtraction of $q$, and one
obtains(consistently with the result stated at the beginning of this
section but now via explicit constructions)
 
\qquad \qquad \qquad number of zero modes = $P_T$.

Static configurations can be considered as a limiting case of periodic
ones (infinite period) but not of antiperiodic ones. The Dirac modes in
a monopole background that introduce the $q$-subtraction as a boundary
effect are not relevant for antiperiodic spinors.
\subsection{Spinors in quasiperiodic backgrounds:}
Consider now the case where in (77) the $k$'s are {\it not} all
mutually commensurable. Generalizing the approximation (48) to several
component irrational ratios one may construct (anti)periodic spinor
solutions at each level of rational approximation. As this approximation
is improved $P_T$ (approx) and hence the number of spinor modes will
diverge. There being no exact period for the gauge field one cannot
limit the number of spinor modes by imposing (anti)periodic boundary
conditions as before.
{\it But our previous solutions provide a subset of "typical" ones which
permit a comparative study of time evolution of spinors in periodic
and quasiperiodic backgrounds respectively.}
Consider,for example, the subset( periodic and antiperiodic
respectively for a periodic background) corresponding to
\begin{eqnarray}
H_j^{(+)}&=&\frac{1}{2r}\Bigl(coth\frac{1}{2}
k_j(r+it-ic_j)+coth\frac{1}{2} k_j(r-it+ic_j)\Bigr) \\
H_j^{(-)}&=&\frac{1}{2r}\Bigl(cosech\frac{1}{2}
k_j(r+it-ic_j)+cosech\frac{1}{2} k_j(r-it+ic_j)\Bigr)
\end{eqnarray}
Suppose we start with mutually commnsurable $k$'s and then vary some of
them( infintesimaliy or more) to introduce incommensurability and
quasiperiodicity. $H_j^{(\pm)}$ will continue to give
solutions, normalizable over any rationally approximated period or over
unit time( analogously to (50)). One can express the spinor densities

$$(\Psi_U^{(\pm)})^\dagger\Psi_U^{(\pm)},\quad 
(\Psi_U^{(\pm)})^\dagger \vec{\tau}\Psi_U^{(\pm)}$$ 
in terms of $a_{\pm},b_{\pm}$ and $\Sigma$. One can implement the
constraints due to the harmonic property of $\Sigma$ and the $H$'s and
(for our present case) those due to spherical symmetry. Then it might be
rewarding to follow the time evolution of the densities for different
values(small or some other crucial ones) of $r$, particularly when
there are three or more incommensurable $k$'s. Such a study is, however,
entirely beyond the scope of this paper.

Let us finally note that starting with (76) as $g_0$ and iterating with
real parameters as in (72), namely
$$g_{p+1}=\frac{a_p+g_p}{a_p+g_p^{-1}}$$
the roots of $g_0=1$ will form a subset of those of $g_p=1$. Hence
defining
\begin{equation}
\Sigma_p=\frac {1}{2r}\Biggl(\frac{1+g_p}{1-g_p}+\frac{1+\bar
g_p}{1-\bar g_p}\Biggr) 
\end{equation}
and still retaining the explicit expressions (85),(86) for $H_j^{(\pm)}$
obtained for $\Sigma_0$, one obtains an interesting subset of spinor
modes for the background corresponding to $\Sigma_p$, for both cases
(periodic and quasiperiodic). At the level of $g_0$ the space-time
dependence can already be quite complex. With iterations this will
become much more so. But we will still have explicit solutions whose
evolutions can be studied.

Note that our spinor solutions (except periodic ones for $q=0$) fall off
exponentially for large
$r$. For antiperiodic ones this is more evident. But though the leading
term (for $q=1$) in
$\Bigl(\frac{H_j}{\Sigma}\Bigr)$ is constant for periodic spinors, the
presence of derivatives introduce again exponential damping. For $q=0$
periodic spinor densities fall off as $r^{-4}$. We have already studied
the action density of the gauge fields near the origin(Sec.2). One can
now study our "exponentially confined" fermion densities near the origin
in such backgrounds.
\section{Propagators:}
For gauge field backgrounds considered in the preceding sections the
propagators for massless, isospin $\frac{1}{2}$ fields were given in
[3]. For periodic backgrounds the (anti)periodic propagators were
presented in explicitly summed up, closed forms. {\it Our iterative map
can be implemented in them} through the functions $g(\bar g)$
corresponding to the points $x$ and $y$ of the propagator
$\Delta{(x,y)}$. We recapitulate briefly the results of [3], where other
sources are cited.

Define
$$t=x_0,t^\prime =y_0,\quad r=\sqrt{{\vec{x}}^2}
,r^\prime=\sqrt{{\vec{y}}^2}$$
$$\sigma_r=\frac{\vec{\sigma}.\vec{x}}{r},\quad \sigma_{r^\prime}
=\frac{\vec{\sigma}.\vec{y}}{r^\prime}$$
$$G=\frac{1+g(r+it)}{1-g(r+it)},\quad
G^\prime=\frac{1+g(r^\prime+it^\prime)}{1-g(r^\prime+it^\prime)}$$
$$\Sigma =\frac{1}{2r} (G+\bar{G}),\quad \Sigma^\prime
=\frac{1}{2r^\prime} (G^\prime+\bar{G^\prime})$$
and

$-i2F(r,t;r^\prime,t^\prime) =$

  $$\frac{(1-\sigma_r)(1+\sigma_{r^\prime})}{(t-t^\prime)
+ i(r-r^\prime)}(G-G^\prime) -  
\frac{(1+\sigma_r)(1-\sigma_{r^\prime})}{(t-t^\prime) -
i(r-r^\prime)}(\bar{G}-\bar{G^\prime}) $$
\begin{equation}
   + \frac{(1-\sigma_r)(1-\sigma_{r^\prime})}{(t-t^\prime) +
i(r+r^\prime)}(G+\bar{G^\prime})  -
\frac{(1+\sigma_r)(1+\sigma_{r^\prime})}{(t-t^\prime) -
i(r+r^\prime)}(\bar G+G^\prime)  
\end{equation}

Let  unprimed fields $(A_\mu)$ correspond to $x$ and primed ones
$(A_{\mu}^{\prime})$ to $y$. Let 
\begin{equation}
\tilde{\Delta}(x,y)=\Sigma^{-\frac{1}{2}}\frac{F}{4\pi^2
(x-y)^2}{\Sigma^\prime}^{-\frac{1}{2}}
\end{equation}
and
$$D_\mu=(\partial_{\mu} + iA_\mu)(\partial_{\mu} + iA_\mu)$$

Then
\begin{equation}
-D^2 \tilde{\Delta}(x,y) = \delta^4 (x-y)
\end{equation}
Thus $\tilde{\Delta}$ gives the {\it aperiodic} propagator for massless
scalar fields. For $G(G^\prime)$ periodic in $t(t^\prime)$ with a period
$T$ , say, the {\it periodic} and the {\it antiperiodic} propagators are
respectively defined to be,
\begin{equation}
\Delta_+ (x,y)=\sum_{n=-\infty}^{\infty} \tilde{\Delta} (x_0
+nT,\vec{x};  y_0,\vec{y})
\end{equation}

\begin{equation}
\Delta_- (x,y)=\sum_{n=-\infty}^{\infty} (-1)^n \tilde{\Delta} (x_0
+nT,\vec{x};  y_0,\vec{y})
\end{equation}
Define
\begin{equation}
V_1=(t-t^\prime)+i\mid \vec{x}-\vec{y} \mid,V_2=(t-t^\prime)-i\mid
\vec{x}-\vec{y} \mid
\end{equation}
$$V_3 (\epsilon, \epsilon^\prime)=(t-t^\prime) + i(\epsilon r +
\epsilon^\prime r^\prime),\quad (\epsilon, \epsilon^\prime = \pm 1)$$
and
$$S_+ (\epsilon, \epsilon^\prime)=
\frac{\pi}{T}\frac{\Bigl((V_2-V_3)cot(\frac{\pi}{T}V_1)+{\rm
cyclic}\Bigr)}{(V_1-V_2)(V_2-V_3)(V_3-V_1)}$$
\begin{equation}
S_- (\epsilon, \epsilon^\prime)=
\frac{\pi}{T}\frac{\Bigl((V_2-V_3)cosec(\frac{\pi}{T}V_1)+{\rm
cyclic}\Bigr)}{(V_1-V_2)(V_2-V_3)(V_3-V_1)}
\end{equation}
(The indices $(\epsilon, \epsilon^\prime)$ are implicit in $V_3$ on
the right hand sides.)
Then one obtains [3],
\begin{equation}
\Delta_\pm (x,y) = \frac{i(\Sigma
\Sigma^\prime)^{-\frac{1}{2}}}{8\pi^2}\left\lgroup\matrix{(G-G^\prime)S_\pm
(1,-1) (1-\sigma_r)(1+\sigma_{r^\prime}) \cr -(\bar
G-\bar{G^\prime})S_\pm (-1,1) (1+\sigma_r)(1-\sigma_{r^\prime}) \cr +
(G+\bar{G^\prime})S_\pm (1,1) (1-\sigma_r)(1-\sigma_{r^\prime}) \cr -
(\bar G+G^\prime)S_\pm (-1,-1)
(1+\sigma_r)(1+\sigma_{r^\prime})}\right\rgroup
\end{equation}
The propagator for spinors is obtained now through a standard
prescription [15] as
$$\Bigl(\gamma .D(x) \Delta (x,y) (1+\gamma_5)/2 + \Delta (x,y)
\gamma.\overleftarrow{D}(y)(1-\gamma_5)/2 \Bigr)$$
where $\Delta$ can be $\tilde{\Delta}$,$\Delta_+$ or $\Delta_-$.

All that has been assumed is that $g$( at $x$,$y$ or at any other
point) saitfies the properties listed under (7). Thus $g$ can involve
an arbitrary number of iterations. When $x$ and $y$ are both close to the
origine, $G$ and $G^\prime$ will exhibit strongly and simultaneously
the consequences of the chaotic aspects studied in App. (See the remarks
in Sec.7.) For a quasiperiodic background one can consider  the
aperiodic $\tilde{\Delta}$ or $\Delta_\pm$ for some adequate rational
approximation.
\section{Generalizations:}
Only brief indications will be given below concerning some possible
generalizations.
\subsection{Beyond spherical symmetry for $q=0,1$ :}
One can generalize (66) and (77) as follows. Let
$$A_\mu=\sigma_{\mu\nu}\partial_\nu ln\Sigma$$
where, with $\epsilon=\pm1$,
\begin{equation}
\Sigma=\frac{1}{2}(1+\epsilon)+\sum_{m=1}^M \frac{1}{2r_m}(G_m+\bar
G_m)
\end{equation}
and $G_m (r_m +it)$ is a holomorphic function with
$$r_m=\mid \vec{x}-\vec{x}_m \mid$$
Choosing
\begin{equation}
G_m=\frac{1+g_m}{1-g_m}
\end{equation}
where $g_m$ is now given by (40) with the origin translated to
$\vec{x}_m$, iterations can be introduced independntly for each centre.
One can have a "gas" (dilute or dense) of (quasi)periodic instantons.
In [7] spinors were studied in such a background without iterations.
One can also start by generalizing (76) including shifts of origin,
starting for the m-th term with
\begin{equation}
g_m^{(0)}=\frac {\sum_{l=1}^{n_m} \lambda_{l,m}^2 \Bigl(e^{k_{l,m}
(r_m+it -ic_{l,m})} - 1 \Bigr)^{-1}}{\sum_{l=1}^{n_m} \lambda_{l,m}^2
\Bigl(1-e^{-k_{l,m} (r_m+it -ic_{l,m})}
 \Bigr)^{-1}}
\end{equation}
Then one may apply iterations independently for each $m$. As explained
in Sec.4 a subset of spinor solutions can readily be obtained.
\subsection{$q>1$; spherical symmetry necessarily broken:}
The ansatz (77) (or (96)) is not suitable for generalizing beyond
$q>1$. This is one of our reasons for starting with (7). The general
formulation of linear pairs [16] can thus be applied specifically [1,4]
to  (quasi)periodic fields for constructing higher $q$ solutions. Details
can be found in those papers. Here we just indicate where iterations
can be implemented.

For $q=2$, one starts with {\it two} functions
\begin{eqnarray}
g_1 &= \prod_{j=1}^n\Biggl({a_j+e^{-k_j(R+it)}\over\bar
a_j+e^{k_j(R+it)}}\Biggr) \nonumber \\
 g_2 &=\prod_{j=1}^n\Biggl({b_j+e^{-k_j(\bar{R}+it)}\over\bar
b_j+e^{k_j(\bar{R}+it)}}\Biggr) 
\end{eqnarray}
where , with real $c$,
$$R=(r^2-c^2-i2cr cos\theta)^{\frac{1}{2}}$$
$$\bar R=(r^2-c^2+i2cr cos\theta)^{\frac{1}{2}}$$
implying an {\it imaginary} translation $(ic)$ parallel to the
$z$-axis. For the solutions to be regular the parameters $(c,a_j,b_j)$
have to satisfy constraints [1,4].
More geneally one starts with $q$ functions $g$ for charge $q$.
Iterations can be implemented for these functions. But one must then
verify carefully the regularity constraints afterwards. That,
presumabbly, would be difficult.
\subsection{Gauge group SU(N):}
Selfdual, (quasi)periodic solutions for arbitrary $N$ were presented
in [5]. Again we only indicate, in the simplest case for $N>2$,
namely $SU(3)$, how the $g$ functions( which can be iterated) appear
in the class of solutions obained. The details of the generalized ansatz
[5] will not be reproduced here.

For $SU(3)$ we just note that, instead of one funtion $e^\zeta$,as in
(5),(6) and (7),one needs two
\begin{eqnarray}
e^{-b_1} &= d_1 P^{-1} (r^2(\partial_r^2 +
\partial_t^2)(g\bar g ))^{-1}\Bigl( (p_3-p_2)(g\bar
g)^{p_1}+{\rm cyclic}\Bigr) 
\nonumber \\  
 e^{-b_2} &= d_2 P^{-1} (r^2(\partial_r^2 +
\partial_t^2)(g\bar 
 g))^{-1}\Bigl( (p_3-p_2)(g\bar g)^{2-p_1}+{\rm cyclic}\Bigr)   
\end{eqnarray}
where $d_1,d_2$ are constants( given in [5]),
$$P=(p_2-p_1)(p_3-p_1)(p_3-p_2)$$
and $(p_1,p_2,p_3)$ are rational numbers satisfying 
$$p_1<p_2<p_3;\quad p_1+p_2+p_3=3$$
To avoid certain problems concerning branch points(explained in [5])
we set   
$$ g = \Biggl(\prod_{j=1}^n\Biggl({a_j+e^{-k_j(r+it)}\over\bar
a_j+e^{k_j(r+it)}}\Biggr)\Biggr)^Q$$
such that $Qp_i,(i=1,2,3)$ are integers.

Iterations can be introduced separately for the factors of $g$ or for
$g$ as a whole. One can also use $g$ of (76) as a starting point.

For higher values of $N$ a set of $N-1$ independent parameters $p$ enters
into the solutions [5]. For $N>2$ a single magnetic winding number is no
longer sufficient for characterizing the asymptotic configurations
in $R^3$. This is one reason for an increasing number of parameters.
\subsection{Use of hyperbolic coordinates:}
The uses of the coordinate transformation
$$(r+it)=tanh\frac{1}{2} (\rho +i\tau)$$
in the construction of instantons or the so called hyperbolic monopoles
were studied in a series of papers.( Apart from [17] Êthey are all
summerized in the review article [18].)
The metric is 
\begin{eqnarray}
ds^2 & = & dt^2+dr^2+r^2 (d\theta^2+(sin\theta)^2 d\phi^2) \nonumber \\
&=& (cosh\rho +cos\tau)^{-2}\bigl(
d\tau^2+d\rho^2+(sinh\rho)^2(d\theta^2+(sin\theta)^2 d\phi^2) \bigr)
\end{eqnarray}
In [17,18] $\tau$-static solutions( depending on $(r,t)$ through
$\rho$) were considered. In [2,4] we indicated the passage from the
$t$-periodic to the $\tau$-periodic solutions. One can replace in (6)
the subscripts $(r,t)$ by $(\rho,\tau)$ respectively and set
\begin{eqnarray}
e^\zeta = \frac{sinh\rho}{(1 - g\bar{g})} \Bigl( ({\partial_{\rho}}^2 + 
{\partial_{\tau}}^2)(g\bar{g})\Bigr)^{ \frac{1}{2}}
\end{eqnarray}
and, for example,
\begin{equation} 
g = \prod_{j=1}^n\Biggl({a_j+e^{-k_j(\rho+i\tau)}\over\bar
a_j+e^{k_j(\rho+i\tau)}}\Biggr)
\end{equation}
The action is evaluated in [2].
To avoid irregularities at
$$\rho=0,\tau =\pm \pi$$
the $k$'s in (103) have to be integers.( The role of the conformal
factor in (101) is crucial concerning this point [18].) This is
a restriction.
 But the formalism is more general in the follwing sense. A
simple rescaling
$$\rho=\lambda r^\prime, \tau=\lambda t^\prime;\quad A_\tau
=\lambda^{-1} A_{t^\prime},A_\rho =\lambda^{-1} A_{r^\prime}$$
gives back the previous formalism, in the limit $\lambda \rightarrow 0$,
with 

$$ d{s^\prime}^2  =4\lambda^{-2} ds^2= 
d{t^\prime}^2+d{r^\prime}^2+{r^\prime}^2 (d\theta^2+(sin\theta)^2
d\phi^2)$$
The restriction on the $k$'s can be lifted after the scaling limit
is taken.

But let us consider the situation without any such rescaling. As it
stands, $g$ in (103) has already a quite special type of $(r,t)$
dependence via $(\rho,\tau)$. After several iterations one can have a
very complex $(r,t)$ dependence, say for the action density. But the
situation can still be studied relatively simply using $(\rho,\tau)$.

The coordinate transformation introduced maps the $(r,t)$ half-plane on
the strip 
$$0\leq \rho < \infty,\quad  -\pi \leq \tau \leq \pi .$$
More generally one can consider our solutions in the context of three
non-compact and one compactified dimension, all with the same signature.
If $T$ is the period associated with the last one then the condition
concerning integer values, say $n_j$ of $k_j$ is to be generalized to
$k_j=(2\pi n_j)/T$ assuring single valued solutions. One can also
consider the possibility of embedding our solutions into such subspaces
of higher dimensional spaces.

\section{Remarks:}
Chaos in gauge theories is a popular topic. Many authors have studied
various aspects of this domain. A convenient reference is [19], a book
devoted to this field with a long bibliography. Among more recent papers
 one may note [20] again citing many sources. Compared to most of the
above-mentioned studies, ours is more modest in one respect but more
ambitious in another.

We have shown (App) that our iterative map, though simple, is a
chaotic one. But it's implementation upto any given order does not
automatically render our field configurations fully chaotic. The
precise way in which the implemented iterations influence the time
evolution of the gauge field has been pointed out in Sec.1. When the
configuration is strictly periodic it comes back, by definition, to it's
initial state after each period $T$. But the sensitive time dependence
discussed in Sec.1 implies that any quantity (such as the action density
at a given point or within a small volume) can fluctuate, within a
single period, more and more with higher number of iterations and in
a more complex fashion. It can wander far and waywardly before coming
back. When the configuration is quasiperiodic it does not return
repeatedly to an inital state (Sec.3) though arbitrarily close
approaches are possible. How should one precisely characterize such
fields with increasingly sensitive time dependence generated by
iterations ? We have not adequately explored the implications, the
consequences of the two most striking features of our solutions, namely
sensitive time dependence and jumps in rotation numbers. We have
exhibited their existence. A more thorough exploration can probably
indicate a satisfactory characterization. At this stage, after a "large"
number of iterations, "at the edge of chaos" might be a convenient
description (though precaution is necessary due to broad,
fashionable uses of this terminology). Sufficiently accurate numerical
studies can help in understanding. But that is beyond the competence of
the present author.

Having noted the limitations, one may now note the positive qualities
of our approach. 
In order to be able to apply techniques develped for one dimentional
dynamical systems authors frequently consider gauge fields depending,
effectively, on one coordinate only. Thus, for example, one studies
time evolutions of fields homogeneous in space ([21],[20] and a
number of papers cited in [19]). In [22] the static problem is
reduced to the one dimensional Duffing equation and then time
dependence is introduced as a perturbation. It is known that,
considred in full generality, Yang - Mills fields are non-integrable
([19] and sources cited there). The fully chaotic aspects are then
related to this non-integrability.

Our approach, one may say, is antithetic to the preceding one. The
intriguing features we exhibit via the mapping arise in fully
integrable (explicitly solved) selfdual configurations. We start
with solutions, found in our previous papers, which have a whole
range of remarkable properties, quite apart from those revealed in
the present study. They combine topological aspects of standard
instantons and monopoles and are charactrized by two topological
numbers (both integers for periodic solutions).Instead of
considering solitonic and chaotic aspects to be entirely
incompatible (and leaving it at that) we have tried to explore,
using our mapping, how close and with what possible limitations,
they can be brought together. This has revealed probably
hitherto unsuspected possibilities. 

Quark confinement has been studied using approaches as
different as that of a dual superconductor and that of random
fields (see Ch.11 of [19]). Our spinors (except for one
subclass) provide explicit solutions damped exponentially away
from the origin. This is not confinement but worth noting. It
should also be noted that this damping {\it increases} with
the temperature (with some typical frequency $K$, inverse of
the period), while confinement is usually supposed to break
down at a sufficiently high temperature. Spinors in
backgrounds with several incommensurable frequencies should
be further explored for a better understanding. Rather than
being fully, dully chaotic they might provide a terrain
fevourable to genesis of rewardingly complex patterns and
sructures. 

Periodic instantons have been considered from the beginning [23,24,25]
to be of particular interest in the study of gauge fields at finite
temperatures. Our more general solutions (as noted in Sec.1) show very
clearly that strict periodicity involves extremely fine tuning of a set
of paramters (the $k$'s). Irrationals being dense on the real line,
infinitesimal shifts in one or more $k$'s can make a periodic solution
quasiperiodic and vice versa. Since such solutions exist it would be
quite artificial to consider periodic solutions exclusively.
Numerically, as is well-known, it is a delicate task to distinguish
between commensurable and incommensurable cases. But since their
mathematical properties are strikingly different, can one understand
better the significance, the role of quasiperiodicity in the context of
finite temperature ? In a realistic situation the temperature cannot
be absolutely steady. A suitably chosen interval covering small
fluctuations of a roughly steady temperature would cover a continuum of
frequencies. Slowly varying temperatures would again need different
considerations. Though such aspects might be worth considering, we have
no simple adequate answer concerning the role of quasiperiodicity in the
context of finite temperatures. We have studied quasiperiodicity for the
light they shed on topological aspects and the possibilities formally
engendered by the presence of several incommensurable periods in the
background.

We have obtained the propagators for (quasi)periodic backgrounds in
particularly convenient forms, where one can implement an arbitrary
number of our iterations. One can next try to compute the fluctuation
determinants. Then one can start to carry over the consequences of the
chaotic features of our mapping into quantum domains through
semi-classical developments. It would be interesting to see in what
fashion and to what extent such features can thus seep through.

Our spinor solutuions are limited to zero mass and to the gauge group
$SU(2)$. But they, along with the propagator for spinor fields (Sec.5),
can provide a starting point, exploiting the full range of our selfdual
solutions, for the study of quarks in a finite temperature gluon
background.
\\[1cm]

 I thank Pierre Collet for discussions concerning iterative maps.

\newpage
\begin{center}
\section*{Appendix}
\end{center}
\setcounter{equation}{0}
\renewcommand{\theequation}{A.\arabic{equation}}

An iterative map of the unit disc is presented and some of it's
properties are studied. In Sec.1 we indicate how this mapping is
implemented in the construction of (quasi)periodic gauge fields and with
what consequences. See also the remarks in Sec.7.
\setcounter{subsection}{0}
\renewcommand{\thesubsection}{A.\arabic{subsection}}
\subsection{The map:}
Let $Z_p$ be a point in the unit disc, centered at the origin, in the
complex plane. Consider the map, with $0<\mid a_p\mid <1$,
\begin{equation}
Z_{p+1}=\frac{a_p + Z_p}{\bar a_p + Z_p^{-1}} = Z_p\frac{a_p + Z_p}{1
 + \bar a_p Z_p}
\end{equation}
For $0 \leq \mid Z_p \mid \leq 1$ one obtains $0 \leq \mid Z_{p+1} \mid
\leq 1$.

The inverse power in the denominator $(Z_p^{-1})$ leads to properties
quit different from those of the standard M\"obius type maps. The most
evident difference is that here zero is a fixed point. But there are
other profound differences. To take just one example, the Schwarzian
derivative (identically zero for M\"obius maps) makes an interesting
contribution in the action density near the spatial origin( Sec.2).
One may set 
$$a_p=a_{p-1}= \cdots =a_0 =a$$
A more general possibility is
\begin{equation} 
a_p = f(a_{p-1})
\end{equation}
with $f$ so chosen as to guarantee $0<\mid a_p\mid <1$ for $0<\mid
a_0\mid <1$. An interesting example of $f$ will be given below. But the
explicit form of $f$ will usually be unspecified, leaving room for
eventual different convenient choices.

The crucial role of a suitable $(r,t)$ parametrization is emphasized in
Sec.1. For the choice (2), $\psi_0$ being the phase of $Z_0$,
\begin{equation}
Z_0 = \mid Z_0 \mid e^{-i\psi_0} = e^{-k(r+it)}; \quad (k>0,r\geq
0,-\infty <t<\infty)
\end{equation}

For the gauge fields we are finally interested in the space-time
dependence. {\it But in App $(r,t)$ should be considered as convenient
notations} defining
\begin{equation}
ln\mid Z_0 \mid=-kr,\quad \psi_0 = kt
\end{equation}

If (3) is chosen for $Z_0$, some different notation, say
$(\widehat{kr},\widehat{kt})$, would be more appropriate. We will
however continue to use in App $(r,t)$ as in (A.4), hoping that no
confusion can arise. Moreover some of our comments will refer
specifically to (2) or (A.3).

Define
$$Z_p = \mid Z_p \mid e^{-i\psi_p}, a_p = \mid a_p \mid e^{-i\alpha_p},
\chi_p=\psi_p -\alpha_p$$
Then from (A.1),
\begin{equation}
\mid Z_{p+1} \mid e^{-i(\psi_{p+1} - 2\alpha_p)} = \mid Z_p \mid
e^{-i\chi_p}\frac{\mid a_p \mid +\mid Z_p \mid
e^{-i\chi_p}}{1 +\mid a_p \mid \mid Z_p \mid
e^{-i\chi_p}}
\end{equation}

\subsection{Moduli:}
One has 
\begin{equation} 
\mid Z_{p+1} \mid^2  =  \mid Z_p \mid^2 \frac{\mid a_p \mid^2 +\mid Z_p
\mid^2 + 2\mid a_p\mid \mid Z_p \mid cos\chi_p}{1 +\mid a_p \mid^2 \mid
Z_p\mid^2 + 2\mid a_p\mid \mid Z_p \mid cos\chi_p} 
\end{equation}
$$ \equiv  \mid Z_p\mid^2 \mid {\tilde{Z_p}}\mid^2$$

where
\begin{equation} 
(1 - \mid {\tilde{ Z_p}} \mid^2 ) =   \frac{(1 -\mid a_p \mid^2)
(1 - \mid Z_p\mid^2 )}{1 +\mid a_p\mid^2 \mid Z_p\mid^2 + 2\mid a_p\mid
\mid Z_p \mid cos\chi_p}\quad \geq 0 
\end{equation}

Thus for
$$\mid Z_p \mid = 0,1, <1;\quad \mid Z_{p+1} \mid =0,1,< \mid Z_p \mid $$
respectively.

Hence not only is zero a fixed point, but the circumference of the disc 
($\mid Z_0 \mid = 1$ or $r=0$) is stable as a whole leading to circle
maps to be studied in detail soon.
For the inverse map, both the roots of the quadratic
\begin{equation}
Z_P^2 + (a_p -  \bar a_p Z_{p+1})Z_p = Z_{p+1}
\end{equation}
must correspond to $0<\mid Z_p \mid <1$ for  $0<\mid Z_{p+1} \mid <1$.
( For $ \mid Z_p\mid \geq 1$,$\mid {\tilde{Z_p}} \mid \geq 1$ and hence
$\mid Z_{p+1}\mid \geq 1$.) These roots coincide for 
\begin{equation}
Z_{p+1} = - \Biggl(\frac{a_p}{\bar a_p}\Biggr) \Biggl(\frac{1
-\sqrt{1-\mid a_p\mid^2}}{1 +\sqrt{1-\mid a_p\mid^2}}\Biggr)
\end{equation}
to
\begin{equation}
Z_p = - \Biggl(\frac{a_p}{1 +\sqrt{1-\mid a_p\mid^2}}\Biggr)
\end{equation}
Thus starting with $\mid Z_0 \mid <1$ ( or $r>0$), $\mid Z \mid $ moves
away under iterations towards the attractive fixed point ($ Z =0$).
Generically two values of $Z_p$ (which may coincide) are mapped on a
$Z_{p+1}$ with a lesser modulus. But instead of a stepwise migration
towards the fixed point zero, the latter might be reached abruptly due
to the following feature.
\subsection{Zeros:}
One has $Z_{p+1}=0$ for
$$ Z_p= 0$$
and for  
$$Z_p= -a_p.$$

The second possibility is worth further study. As will be seen later,
the more general condition $\mid Z_p\mid = \mid a_p\mid$ provides
crucial domains of discontinuities of rotation numbers associated to the
phases. Moreover
$$Z_p= Z_{p-1}\frac{a_{p-1} + Z_{p-1}}{1+ \bar a_{p-1}Z_{p-1}} = -a_p$$
furnishes an interesting context for exploring the consequences of
different choices for $f$ in $a_p=f(a_{p-1})$. One has
\begin{equation}
\Biggl(\frac{Z_{p-1}}{\sqrt a_p}\Biggr)^2 + 2\lambda_p
\Biggl(\frac{Z_{p-1}}{\sqrt a_p}\Biggr) +1 =0
\end{equation}
where 
$$\lambda_p = \bigl(2\sqrt a_p \bigr)^{-1} \bigl(a_{p-1} + a_p \bar
a_{p-1}\bigr)$$
Hence 
$$Z_{p-1}=\sqrt a_p \Bigl( -\lambda_p \pm \sqrt{\lambda_p^2 -1}\Bigr)$$

The two roots coincide to $Z_{p-1}=\mp \sqrt a_p$ for
$\lambda_p =\pm 1$ or
\begin{equation}
\sqrt a_p=\pm {\Bigl(\bar a_{p-1} \Bigr)}^{-1} \Bigl(1 - \sqrt{1-\mid
a_{p-1} \mid ^2} \Bigr)
\end{equation}
Denoting, with real $\mu_p$,
\begin{equation}
a_p=\bigl(tanh\mu_p \bigr) e^{-i\alpha_p}
\end{equation}
(A.12) gives 
$$ tanh\mu_p = \Bigl(tanh \frac{1}{2}\mu_{p-1}\Bigr)^2 ,\qquad \alpha_p
=2\alpha_{p-1}$$
This is an example of the choice of $f$ assuring special properties 
(here double zeros). For comparison note that choosing , for all $p$,
$$a_p = \bar a_p = a, \qquad (0<a<1)$$
\begin{equation}
\lambda_p =\frac{1}{2} \sqrt a (1+a) = cos\zeta
\end{equation}
say, $\zeta$ being real. Now (A.11) gives
\begin{equation}
Z_{p-1}=-\sqrt a e^{\mp i\zeta}
\end{equation}
\subsection{Circle map on the circumference (chaotic aspects):}
For $\mid Z_0\mid = 1, (i.e.\quad r=0)$, for all $p$,
$$\mid Z_p\mid = 1.$$

For the phases $\psi_p$, using the notations of (A.5), one has the
circle map

\begin{equation}
e^{-i\psi_{p+1}} = \frac{a_p + e^{-i\psi_p}}{\bar a_p + e^{i\psi_p}} =
e^{-i2\alpha_p}
\frac{\mid a_p\mid + e^{-i\chi_p}}{\mid a_p\mid +e^{i\chi_p}}
\end{equation}

Hence
\begin{eqnarray}
\frac {d\psi_{p+1}}{d\psi_p} & = & 2\Biggl( \frac{1+\mid a_p \mid
cos\chi_p} {1+{\mid a_p \mid}^2 + 2\mid a_p \mid
cos\chi_p}\Biggr) \nonumber\\
&= & 1+ \frac {\bigl(1-\mid a_p \mid^2 \bigr)}{\bigl(1-\mid a_p \mid
\bigr)^2 + 4\mid a_p \mid \bigl(cos(\chi_p /2)\bigr)^2} \quad >1
\end{eqnarray}

Thus $\psi_{p+1}$ is {\it monotonic} (increasing) in $\psi_p$ and
\begin{eqnarray}
e^{-i(\psi_{p+1} -2\alpha_p)} & = & 1 \qquad {\rm for} \quad \chi_p
=0,\pi, 2\pi  \\
&=&  -1 \qquad {\rm for} \quad cos\chi_p = -\mid a_p \mid,\quad
sin\chi_p = \pm \sqrt {1-\mid a_p \mid^2} \nonumber
\end{eqnarray}

Hence as
\begin{eqnarray} 
\psi_p &\rightarrow &\quad \psi_p +2\pi \nonumber \\
\psi_{p+1} &\rightarrow & \quad \psi_{p+1} +4\pi
\end{eqnarray}

(This result also follows from an approach analogous to the one leading
to (A.52) for $\mid Z_0\mid >\quad  \mid a_0\mid$, since here $\mid
Z_p\mid =1 >\quad \mid a_p\mid$. But the foregoing instructive one
follows the rotations in more detail.) The result (A.19) is fudumental
in computing the actions of the gauge fields after iterations (Sec.2).
But here we concentrate on another aspect.

It is well-known ([9], p.50 and also p.18) that the apparently very
simple circle map 
\begin{equation}
\theta_{p+1} = 2\theta_p, \qquad \frac{d\theta_{p+1}}{d\theta_p} =2
\end{equation}
is {\it chaotic}. It satisfies all the requisite conditions, the most
important one being a srtong sensitiveness to initial conditions (here
in the form of expansiveness with index $ln2$). Our example (A.16) will
be seen to satisfy the same conditions, but in a subtler fashion. In
fact, our case contains (A.20) as a particularly simple limit $(a_p =0)$.
"On the average" $\psi_{p+1}$ turns twice as fast as $\psi_p$. But the
rate is less than twice in one domain and just sufficiently more than
twice in the complementary one to compensate.

For
\begin{eqnarray}
cos\chi_p &= &\pm 1 \nonumber \\ 
\quad v_p \equiv \frac{d\psi_{p+1}}{d\psi_p}
&=&\frac{2}{1 \pm \mid a_p \mid}
\end{eqnarray}
and for 
$$ cos\chi_p = - \mid a_p \mid, \quad v_p = 2$$

The complementary domains are
$$1\geq cos\chi_p > -\mid a_p \mid \qquad (v_p <2)$$
$$-1\leq cos\chi_p < -\mid a_p \mid \qquad (v_p >2)$$
As $\mid a_p \mid \rightarrow 1$ the first domain increases, but so does
$v_p$ in the other to compensate (becoming very high near $cos\chi_p
=-1$).

After $p$ iterations (with $\psi_0 = kt$)
\begin{eqnarray}
\frac{d\psi_p}{d\psi_0}&=&\frac{1}{k}\frac{d\psi_p}{dt}=\prod_{j=0}^{p-1}
v_j \nonumber \\
&=& \prod_{j=0}^{p-1}\Biggl( \frac{2\bigl(1+\mid a_j\mid cos\chi_j
\bigr)}{1+ \mid a_j\mid^2 + 2\mid a_j\mid cos\chi_j}\Biggr)
\end{eqnarray}
and
\begin{eqnarray}
\frac{d^2\psi_p}{d\psi_0^2} = \frac{1}{k^2}
\frac{d^2\psi_p}{dt^2}=\frac{d\psi_p}{d\psi_0}\Biggl(\sum_{j=0}^{p-1}V_j\Bigl(\frac{d\psi_j}{d\psi_0}\Bigr)\Biggr)
\end{eqnarray}
where
$$V_j = \frac{\bigl(1-\mid a_j\mid^2\bigr)\mid a_j\mid
sin\chi_j}{\bigl(1+\mid a_j\mid cos\chi_j
\bigr)\bigl(1+\mid a_j\mid^2 + 2\mid a_j\mid cos\chi_j\bigr)}$$

In (A.22) each factor $v_j  >1$. But due to the factor $sin\chi_j$ in
$V_j$ the terms in (A.23) can change sign.

The sensitive dependence on initial data should be evident from the
preceding analysis. But let us formulate it more precisely in terms of a
{\it characteristic index}. For (A.20) the index is evidently $ln2$ [9].
This is recovered in our case in the limit of each $a_j =0$. For (A.16)
it depends on the sequence $(a_0,a_1,\ldots )$. {\it But it has a
positive lower bound}. In (A.22) replacing each $v_j$ by it's upper and
lower bound respectively
\begin{equation}
\prod_{j=0}^{N-1} \Bigl(\frac{2}{1-\mid a_j\mid}\Bigr) \geq
\frac{d\psi_N}{d\psi_0}\geq \prod_{j=0}^{N-1} \Bigl(\frac{2}{1+\mid
a_j\mid}\Bigr) >1
\end{equation}
Setting for simplicity all $\mid a_j\mid =\mid a\mid$, the
characteristic index $\lambda$  satisfies
\begin{equation}
ln \Bigl(\frac{2}{1-\mid a\mid}\Bigr) \geq \lambda \geq ln
\Bigl(\frac{2}{1+\mid a\mid}\Bigr) > 0
\end{equation}
More generally $ \Bigl(\frac{2}{1 \pm \mid a\mid}\Bigr)$ should be
considered as the geometric means of the corresponding products in
(A.24).

For a map to be chaotic it must have a dense set of periodic points [9].
For (A.20) the periodic points are given by [9]
\begin{equation}
\theta_n = 2^n \theta = \theta + 2k\pi \quad {\rm or} \quad \theta =
\frac{2k\pi}{2^n -1}
\end{equation}
where $k$ is an integer and
$$0\leq k\leq 2^n -1$$
This is the situation for all $a$'s zero in (A.16). But for (A.16) one
cannot obtain as simply a general formula. One can however proceed
stepwise to show how the periodic points remain dense but are shifted
as the parameters $a$ incrase from zero. For simplicity consider all
$a_p$'s equal and real. Then
$$e^{-i\psi_{p+1}} = \frac {a+e^{-i\psi_p}}{a+e^{i\psi_p}}$$
gives
\begin{equation}
\psi_{p+1} = 2\psi_p -2\bigl( a sin\psi_p - \frac{1}{2} a^2sin2\psi_p
+\ldots \bigr)
\end{equation}
Thus 
\begin{equation}
\psi_n = 2^n \psi_0 - a \Bigl(\sum_{l=1}^n 2^l sin2^{n-l} \psi_0 \Bigr)
+{\rm O}(a^2)
\end{equation}

Hence upto O($a$) the periodic points are given by 
\begin{equation}
\psi =\frac{2k\pi}{2^n -1} + aS_1
\end{equation}
where
$$S_1 = \Bigl(\frac{2^n}{2^n -1}\Bigr)\Biggl(\sum_{l=1}^n 2^{-n+l}
sin2^{n-l}\Bigl(\frac{2k\pi}{2^n -1}\Bigr)\Biggr)$$
$$ < \Bigl(\frac{2^n}{2^n -1}\Bigr)\Bigl(\sum_{l=1}^n 2^{-n+l}\Bigr) =2$$
Thus the dense set (A.26) is shifted as shown above. One may now iterate
to higher powers of $a$ and find an analogous situation. We cannot
produce a general solution for (A.16) in a closed form, but the smooth
continuity with (A.26) is clear enough.

A third criterion for chaoticity [9] is topological transitivity. This
is satisfied by (A.16) as obviously as by (A.20). Our preceding analysis
of rotations makes it evident that their effects can not remain confined
in one particular segment of a circle.

{\it Thus all the three criteria for being chaotic are satisfied by our
map.}

\subsection{Series expansion near the circumference (small $r$):}

For the gauge field configurations principally considered in this paper
the time dependence is exponentially damped as $r$ increases. So the
(quasi)periodic time evolution is best studied for small $r$. For our
mapping this corresponds to a domain near the cicumference at a distance 
$(1-e^{-kr})$.

Let $\psi_p$ {\it continue} to denote the value of the phase for $r=0$
and let
\begin{equation}
Z_p = e^{-i\psi_p} \bigl( 1+ C_p^{(1)} r +C_p^{(2)} r^2 + \ldots )
\end{equation}
where the $C$'s can be {\it complex} since the $r$-dependence of the
total phase is included in them. The expansion (A.30) is general, but
particularly useful for small $r$ due to evident reasons. (
Exceptionally, in this subsection only, $\psi_p$ denotes not the total
phase but a part. The notation $\psi_p^{(0)}$ would have been more
consistent. But this simplification, leaving room for other indices to
come, should not cause confusion.)

{\it Suppresing the index} $p$ {\it temporarily} and using the
holomorphy condition
\begin{equation}
\bigl(\partial_r + i\partial_t \bigr)Z = 0
\end{equation}
one obtains
\begin{equation}
C^{(1)} +2C^{(2)}r+3C^{(3)}r^2+\dots  = -\frac{d\psi}{dt}
\bigl(1+C^{(1)}r + C^{(2)}r^2 +\dots \bigr)  
 -i\bigl(\frac{dC^{(1)}}{dt} r + \frac{dC^{(2)}}{dt}r^2 + \dots \bigr)
\end{equation}
Thus 
$$C^{(1)} = -\frac{d\psi}{dt}$$
and, for $l>1$,
\begin{equation}
lC^{(l)} = C^{(1)}C^{(l-1)}  -i\frac{dC^{(l-1)}}{dt}
\end{equation}
The general solution is
\begin{equation}
C^{(l)} = \frac{e^{i\psi}}{l!}\Bigl( -i\frac{d}{dt}
\Bigr)^l\Bigl(e^{-i\psi}\Bigr)
\end{equation}

{\it This expansion displays precisely how the chaotic properties of the
circle map for the phase are carried over through  $\psi$  and its
derivatives in the coefficients.}

The result (A.34) is compact and elegant. But separate explicit
expressions for the total phase and the amplitude of $Z$ are useful for
the gauge fields. We give the first few tems of the $r$-expansion for
both. The coefficients $(\psi^{(l)}, B_l)$ will now all be real. One
obtains, keeping terms upto O($r^4$),
\begin{eqnarray}
Z&=&e^{-i\psi}\bigl( 1+C_1 r+C_2 r^2+C_3 r^3+C_4 r^4\bigr) \nonumber\\
&=& e^{-i(\psi^{(0)} +\psi^{(2)}r^2 +\psi^{(4)}r^4)}\Bigl( 1+B_1 r +B_2
r^2 + B_3 r^3 + B_4 r^4 \Bigr)
\end{eqnarray}
The coefficients $\psi^{(1)}$ and $\psi^{(3)}$ turn out to be zero and
one obtains, in terms of the $C$'s given before,
\begin{eqnarray}
\psi^{(2)} &=& \frac{i}{2}\bigl( C^{(2)} - \bar C^{(2)} \bigr)
\nonumber\\
\psi^{(4)} &=& \frac{i}{2}\Bigl(\bigl( C^{(4)} - \bar C^{(4)}
\bigr)-\frac{1}{2}\bigl( (C^{(2)})^2 - (\bar C^{(2)})^2 \bigr)\Bigr)
\end{eqnarray}
and
\begin{eqnarray}
B_1 &=& C^{(1)} \nonumber \\
B_2 &=& \frac{1}{2}\bigl( C^{(2)} +\bar C^{(2)} \bigr) \nonumber \\
B_3 &=& \frac{1}{2}\bigl( C^{(3)} +\bar C^{(3)} \bigr) \nonumber \\
B_4 &=& \frac{1}{2}\bigl( C^{(4)} +\bar C^{(4)} \bigr)
-\frac{1}{8}\Bigl( (C^{(2)})^2 + (\bar C^{(2)})^2 \Bigr)
\end{eqnarray}
One may also note that
\begin{equation}
Z\bar Z =  1+D_1 r+D_2 r^2+D_3 r^3+D_4 r^4 + {\rm O}(r^5)
\end{equation}
where
\begin{eqnarray}
D_1&=&2C^{(1)} \nonumber \\
D_2&=&2\Bigl(C^{(1)}\Bigr)^2 \nonumber \\
D_3&=&\frac{4}{3}\Bigl(C^{(1)}\Bigr)^3 -
\frac{1}{3}\frac{d^2}{dt^2}C^{(1)}\nonumber\\
D_4&=&\frac{2}{3}\Biggl(\Bigl(C^{(1)}\Bigr)^4  -
C^{(1)}\frac{d^2}{dt^2}C^{(1)}\Biggr)  
\end{eqnarray}
Higher order terms can be evaluated stepwise. These results hold, of
course, for any $p$. If the product of several factors $\bigl(Z_p
{Z^{\prime}}_{p^\prime} \ldots \bigr)$ is considered, one has the same
expansion with
\begin{equation}
C^{(1)}= -\frac{d}{dt} \Bigl( \psi_p + {\psi^\prime}_{p^\prime} +\ldots
\Bigr)
\end{equation}
Consistency with this constraint is a useful check on the numerical
coefficients obtained above. The notation indicates that for each factor
the sequence of the parameters $a$, the periods involved and also the
order of iterations can be different. The results (A.39) yield the
leading term in the action density near the spatial origin (Sec.2).
\subsection{Annular maps and rotation numbers:}
We have studied some interesting proprties of our map on the
circumference of the unit disc and nearby $(r=0,r \ll 1)$. One can
continue an analogous study away from the edge. But we now concentrate
on a different class of remarkable features associated to specific
values of $r$ as it increases.

For $\mid Z\mid =1$ the iterations affect only the phase giving a circle
map. For $\mid Z\mid <1$ the amplitude also changes. It diminishes and
becomes a function of the phases of the previous steps. The domain of
variation of $Z$ (as a function of these phases) becomes an annulus,
which can, crucially, become a disc. To emphasize this aspect we use the
term "annular map". The rotation numbers to be defined will be associated
to the phases. We are fundumentally interested in variations of functions
of the phase of $Z_0$ as the latter moves on a circle of radius
$e^{-kr}$. this provides, most directly through (2), the link with the
time evolution of gauge fields.

From (A.5) and (A.6), with the notations defined there $(\chi_p =\psi_p
-\alpha_p, \ldots )$
\begin{equation} 
\mid Z_{p+1} \mid^2  =  \mid Z_p \mid^2 \frac{\mid a_p \mid^2 +\mid Z_p
\mid^2 + 2\mid a_p\mid \mid Z_p \mid cos\chi_p}{1 +\mid a_p \mid^2 \mid
Z_p\mid^2 + 2\mid a_p\mid \mid Z_p \mid cos\chi_p} 
\end{equation}
and
\begin{equation}
\psi_{p+1} = \psi_p +\frac{i}{2}\bigl( ln f_1 + ln f_2 \bigr) +2\alpha_p
\end{equation}
where
$$ f_1 = \Biggl( \frac{\mid a_p \mid + \mid Z_p \mid e^{-i\chi_p}}{\mid
a_p\mid + \mid Z_p\mid e^{i\chi_p}}\Biggr),\quad f_2 =
\Biggl(\frac{1+\mid a_p\mid \mid Z_p\mid e^{i\chi_p}}
{1+\mid a_p\mid \mid Z_p\mid e^{-i\chi_p}}\Biggr)$$
In particular, for $p=0$ with $Z_0=e^{-k(r+it)}$,
\begin{equation} 
\mid Z_{1} \mid^2  =  e^{-2kr} \frac{\mid a_0 \mid^2 +e^{-2kr} + 2\mid
a_0\mid e^{-kr} cos\chi_0}{1 +\mid a_0 \mid^2 e^{-2kr} +
2\mid a_0\mid e^{-kr} cos\chi_0} 
\end{equation}
For fixed $r$, this can be shown to be a monotonic increasing function
of $cos\chi_0$. One obtains
\begin{equation}
{\mid Z_1\mid}_{max}={\mid Z_0\mid}\Biggl(\frac{\mid a_0\mid+\mid
Z_0\mid}{1+\mid a_0\mid\mid Z_0\mid}\Biggr)
\end{equation}
\begin{equation}
{\mid Z_1\mid}_{min}={\mid Z_0\mid}\Biggl(\frac{\mid a_0\mid-\mid
Z_0\mid}{1-\mid a_0\mid\mid Z_0\mid}\Biggr)
\end{equation}
The width of the annulus is
$$W_1={\mid Z_1\mid}_{max}-{\mid Z_1\mid}_{min}$$
For $\mid Z_0\mid>\mid a_0\mid$
\begin{equation}
W_1= 2\mid a_0\mid\mid Z_0\mid \Biggl(\frac{1-\mid
Z_0\mid^2}{1-\bigl(\mid a_0\mid\mid Z_0\mid\bigr)^2}\Biggr)
\end{equation}
For $\mid Z_0\mid<\mid a_0\mid$
\begin{equation}
W_1= 2\mid Z_0\mid^2 \Biggl(\frac{1-\mid
a_0\mid^2}{1-\bigl(\mid a_0\mid\mid Z_0\mid\bigr)^2}\Biggr)
\end{equation}
For$\mid Z_0\mid=\mid a_0\mid$, {\it the annulus becomes a disc} with
$${\mid Z_1\mid}_{min}=0$$ 
and
\begin{equation}
W_1= \frac{2\mid a_0\mid^2}{1+ \mid a_0\mid^2}
\end{equation}
As $\mid Z_1\mid$ touches zero the phase of $Z_1$ becomes undefined
with crucial consequences. We will study them now.

Our aim is to compare the rates of rotation of $\psi_0$ and $\psi_1$ for
the following domains of $\mid Z_0\mid = e^{-kr}$,
$$\mid Z_0\mid >\mid a_0\mid,$$
$$\mid Z_0\mid =\mid a_0\mid$$
and
$$\mid Z_0\mid  <\mid a_0\mid$$

For $\mid Z_p\mid =1(> \mid a_p\mid)$ we have already seen ((A.16) to
(A.21)) that on the average $\psi_{p+1}$ turns twice as fast as
$\psi_p$. In particular,$\psi_1$ turns twice as fast as $\psi_0$. This
result will be seen to hold more genrally for $\mid Z_0\mid >\mid
a_0\mid$. But discontinuities appear as  $\mid Z_0\mid$ comes down to
$\mid a_0\mid$ and crosses over. We will demonstrate this now.

Let us examine (A.42) for $p=0$ and in particular the term $lnf_1$. For 
$\mid Z_0\mid >\mid a_0\mid$ the real part of $\bigl(\mid a_0\mid+\mid
Z_0\mid e^{-i\chi_0}\bigr)$ can vanish and change sign. {\it This
feature is absent in $f_2$ since always $\mid a_p\mid\mid
Z_p\mid <1$.} So the term $lnf_1$ has to be treated carefully to keep
track of additive contributions as $\psi_0$ rotates ($t$ increases). It
is convenient to proceed as follows.
For
$$ \mid Z_0\mid >\mid a_0\mid$$
\begin{equation}
f_1=e^{-i2\chi_0}\Biggl(\frac{1+\frac{\mid a_0\mid}{\mid
Z_0\mid}e^{i\chi_0}}{1+\frac{\mid a_0\mid}{\mid
Z_0\mid}e^{-i\chi_0}}\Biggr)
\end{equation}
For
$$ \mid Z_0\mid =\mid a_0\mid$$
\begin{equation}
f_1=e^{-i\chi_0}
\end{equation}
For  
$$ \mid Z_0\mid <\mid a_0\mid$$
\begin{equation}
f_1=\Biggl(\frac{1+\frac{\mid Z_0\mid}{\mid
a_0\mid}e^{-i\chi_0}}{1+\frac{\mid Z_0\mid}{\mid
a_0\mid}e^{i\chi_0}}\Biggr)
\end{equation}
 
Hence for the three domains, respectively,
\begin{eqnarray}
\psi_1 &=& 2\psi_0 + \Lambda_{(+)} \nonumber \\
 \psi_1 &=&\frac{3}{2}\psi_0 + \Lambda_{(0)} \nonumber \\
 \psi_1 &=& \psi_0 + \Lambda_{(-)}
\end{eqnarray}
where one can now safely consider that
\begin{equation}
\Lambda_{\delta} \bigl( \psi_0 +2\pi \bigr) =\Lambda_{\delta} \bigl(
\psi_0  \bigr), \qquad (\delta = +, 0, -)
\end{equation}

Hence for $\psi_0 \rightarrow \psi_0 +2n\pi$ there are no cumulative,
additive contributions from $\Lambda_{\delta}$ giving a supplementary
term proportional to $n$ (as does the first term proportional to
$\psi_0$). The discontinuity involved for $\psi_1$ at $\mid Z_0\mid =\mid
a_0\mid$ is now explicit.

In view of the crucial role of this result we present an alternative
approach, closely following the argument for the circle map ((A.18) to
(A.21)) but generalizing it for all $\mid Z_0\mid >\mid a_0\mid$. (This
also generalizes the  arguments for chaoticity on the circumference for
the interior of the disc.)

It is sufficient to consider one single term $(f_1)$ as follows.
Define
\begin{equation}
e^{-i\beta} = \frac{\mid a_0\mid +\mid Z_0\mid e^{-i\chi_0}}
{\mid a_0\mid +\mid Z_0\mid e^{i\chi_0}}
\end{equation}
when
$$\frac{d\beta}{d\psi_0}=\frac{2\mid Z_0\mid(\mid Z_0\mid+
\mid a_0\mid cos\chi_0)}{(\mid a_0\mid^2+\mid Z_0\mid^2+2\mid a_0\mid\mid
Z_0\mid cos\chi_0)}$$
Hence for $\mid Z_0\mid>\mid a_0\mid$ one has strict monotonicity with
\begin{equation}
\frac{d\beta}{d\psi_0}>0
\end{equation}
but {\it not} for $\mid Z_0\mid<\mid a_0\mid$. For the exceptional value
$\mid Z_0\mid=\mid a_0\mid$ one has simply
\begin{equation}
e^{-i\beta} =e^{-i\chi_0}
\end{equation}
For $\mid Z_0\mid >\mid a_0\mid$ along with strict monotonicity one has
\begin{eqnarray}
e^{-i\beta}& = &-1\quad {\rm for}\quad cos\chi_0= -\frac{\mid
a_0\mid}{\mid Z_0\mid}, sin\chi_0 = \pm\sqrt {1- \frac{\mid
a_0\mid^2}{\mid Z_0\mid^2}} \nonumber \\
e^{-i\beta} &=& 1\quad {\rm for}\quad cos\chi_0= 0, \pi, 2\pi
\end{eqnarray}
Hence, as for the circle map, "on the average" $\beta$ turns twice as
fast as $\psi_0$. This corresponds to the factor $2$ in the first
equation of (A.52), namely,
$$\psi_1=2\psi_0 + \Lambda_{(+)}$$
Now suppose we consider the Poincar\'e sections for 
$$t_n = t+nT \qquad (n= 0,1,2, \ldots )$$
with some suitably chosen period $T$.
The "rotation number" for $\psi_0 (=kt)$ is defined in terms of 
$$\psi_0^{(n)} =kt_n$$
as
\begin{equation}
\Omega_0=\Bigl( n^{-1} \bigl( \psi_0^{(n)} - \psi_0 \bigr) \Bigr)_{n
\rightarrow \infty} =kT
\end{equation}
(We reserve the term "winding number" for the magnetic charge $q$ of the
gauge fields.) Here, of course, it would have been natural to set 
$$T=2\pi k^{-1}$$
or, rescaling, obtain an integer (say, unit) value for $\Omega_0$. But
we keep $T$ unspecified here since for the gauge fields there will be
the simultaneous presence of different periodic building blocks each
with it's own period, which may even be mutually incommensurable. (See
Sec.3.)
Now, similarly defining the rotation number $\Omega_1$ for $\psi_1$,
one obtains from (A.52) and(A.53)
\begin{eqnarray}
\Omega_1 &=& 2\Omega_0 \qquad {\rm for}\quad e^{-kr} > \mid a_0\mid
\nonumber \\  
\Omega_1 &=& \frac{3}{2}\Omega_0 \qquad {\rm for}\quad e^{-kr} = \mid
a_0\mid\nonumber \\
\Omega_1 &=& \Omega_0 \qquad {\rm for}\qquad e^{-kr} < \mid a_0\mid
\end{eqnarray}
Hence there is a step discontinuity at $r=-k^{-1}ln\mid a_0\mid$.
Let us further examine how this implies {\it sensitive dependence on
small differences in the value of a parameter}. Let $\psi_1^{(\pm)}$
and $\hat{\psi_1}$ denote the values of $\psi_1$ respectively for
$$\mid Z_0\mid=\mid a_0\mid (1 \pm \varepsilon) \quad {\rm and} \quad 
 \mid a_0\mid$$
Here $\varepsilon \ll 1$. From (A.42) with $p=0$, upto O($\epsilon$), one
obtains
\begin{equation}
\psi_1^{(\pm)} =\hat{\psi_1} \pm \varepsilon \Biggl(
\frac{1}{2}tan\frac{\chi_0}{2} - \frac{\mid a_0\mid^2 sin\chi_0}
{1+\mid a_0\mid^4+2\mid a_0\mid^2 sin\chi_0}\Biggr)
\end{equation}
But such a development is valid if not only $\varepsilon \ll 1$ {\it  but
also} $\varepsilon tan\frac{\chi_0}{2}\ll 1$. Even if one starts with a
suitably small value of the latter, the constraint will be violated as
$\chi_0$ approaches the values $(2N+1)\pi$. Near such values one may
proceed as follows.
The contribution to $-i2(\psi_1^{(+)}-\psi_1^{(-)})$ from  the term
$lnf_1$ in (A.42) is
\begin{equation}
ln\Biggl(\Bigl(\frac{1+e^{i\chi_0}+\varepsilon}{1+e^{-i\chi_0}+
\varepsilon}\Bigr)
\Bigl(\frac{1+e^{-i\chi_0}-\varepsilon}{1+e^{i\chi_0}-\varepsilon}\Bigr)\Biggr)
=ln \Biggl(\frac{cos(\chi_0 /2) -i(\varepsilon /2)sin(\chi_0 /2)}
{cos(\chi_0 /2) +i(\varepsilon /2)sin(\chi_0 /2)} \Biggr)
\end{equation}
For $\chi_0=(2N+1)\pi$ this is no longer proportional to $\varepsilon$
but becomes $ln(-1)$. The additive contibution at each turn gives a
difference $(\psi_1^{(+)}-\psi_1^{(-)})$ consistent with (A.59).

So far we have analysed the effect of the passage $Z_0 \rightarrow Z_1$.
For the next step, write
\begin{equation}
Z_2=\Biggl(\frac{a_1+Z_1}{{\bar a}_1 +Z_1^{-1}}\Biggr)
=Z_0^{-2}\Biggl(\frac{a_0+Z_0}{{\bar a}_0+Z_0^{-1}}\Biggr)
\Biggl(\frac{\mu_+ +Z_0}{{\bar \mu}_++Z_0^{-1}}\Biggr)
\Biggl(\frac{\mu_-+Z_0}{{\bar \mu}_- +Z_0^{-1}}\Biggr)
\end{equation}
where (compare (A.11) and the discussion that follows)
\begin{equation}
\mu_{\pm}=\sqrt a_1 \Bigl( \lambda_0 \pm \sqrt {\lambda_0^2 - 1}\Bigr)
\end{equation}
with
$$\lambda_0 =\frac{a_0 +\bar a_0 a_1}{2\sqrt a_1}$$
One sees that for studying $\Omega_2$ (the rotation number associated
to the phase $\psi_2$ of $Z_2$) one should extend the previous
considerations to {\it three} spherical shells
\begin{equation}
\mid Z_0\mid = \mid a_0\mid,\quad \mid \mu_+\mid,\quad \mid \mu_-\mid 
\end{equation}
where
$$\mid \mu_+\mid \mid \mu_-\mid =\mid a_1\mid$$
One may get special features corresponding to a possible double zero
when 
$\mid \mu_+\mid$ and  $\quad \mid \mu_-\mid$ coincide to $\sqrt {\mid
a_1\mid}$. The number of possibilities continue to increase with each
iteration. Generically the process is systematic, but roots can coincide
leading to special features. The cumulative effects of the jumps give
more and more elaborate staircase like patterns (Sec.2).

Considering $(\mid Z_p \mid, \psi_p)$ as independent variables one may
carry out a similar analysis for the step $Z_p \rightarrow Z_{p+1}$ (for
$p >0$). One then obtains formally, like (A.52) with the $\Lambda$'s
satisfying the criterion (A.53),
\begin{eqnarray}
\psi_{p+1} &=& 2\psi_p + \Lambda_{(+)}^{(p)} \nonumber \\
 \psi_{p+1} &=&\frac{3}{2}\psi_p + \Lambda_{(0)}^{(p)} \nonumber \\
 \psi_{p+1} &=& \psi_p + \Lambda_{(-)}^{(p)}
\end{eqnarray}
respectively for 
$$\mid Z_p\mid > \mid a_p\mid,\quad \mid Z_p\mid = \mid a_p\mid, \quad 
\mid Z_p\mid <\mid a_p\mid .$$
But it should be clearly noted that for $p \geq 1$ all $\mid Z_p\mid$'s
are time dependent even for the choice (2) for $Z_0$. For the choice
(3) even $Z_0$ is time dependent. The formal steps of the  iterations
and their consequences in the context of the mapping do not depend on
the parametrization of $Z_0$ (provided it satifies $Z_0 <1$). But when
the time evolution is studied in the context of the gauge fields much
depends evidently on the initial choice.

\end{document}